\documentclass[18pt,a4paper,onecolumn,fleqn,usenatbib,useAMS,referee]{mnras}

\usepackage{newtxtext,newtxmath}
\UseRawInputEncoding

\usepackage[T1]{fontenc}

\DeclareRobustCommand{\VAN}[3]{#2}
\let\VANthebibliography\thebibliography
\def\thebibliography{\DeclareRobustCommand{\VAN}[3]{##3}\VANthebibliography}

\usepackage{graphicx}	
\usepackage{amsmath}	
\usepackage{multirow,booktabs}
\usepackage{parskip}
\usepackage{multirow}

\title[Short title, max. 45 characters]{Infrared spectroscopy of astrophysical ice analogs. The effect of incident angle and refractive index of the ice in the infrared.}
\title{Infrared spectroscopy of astrophysical ice analogs at oblique angles}

\author[C. Gonz\'{a}lez D\'{\i}az et al.]{
C. Gonz\'{a}lez D\'{\i}az,$^{1}$\thanks{E-mail: cgonzalez@cab.inta-csic.es (CGD)}
H. Carrascosa,$^{1}$
and G. M. Mu\~{n}oz Caro$^{1}$
\\
$^{1}$Centro de Astrobiolog\'{\i}a (CSIC-INTA), Ctra. de Ajalvir, km 4, Torrej\'on de Ardoz, 28850 Madrid, Spain} 

\date{Accepted XXX. Received YYY; in original form ZZZ}

\pubyear{2024}

\begin{document}
\label{firstpage}
\pagerange{\pageref{firstpage}--\pageref{lastpage}}
\maketitle

\begin{abstract}
\label{abstract}    
In astrochemical exploration, infrared (IR) spectroscopy is vital for understanding the composition and structure of ice in various space environments. This article explores the impact of incident angles on IR spectroscopy, focusing on molecular components present in interstellar and circumstellar ice mantles such as CO, CO$_2$, H$_2$O, CH$_3$OH, NH$_3$, CH$_4$, H$_2$S. The experiment involves changing the angle at which the infrared beam hits the surface used for ice deposition. It is important to measure the density of the ice layer accurately, especially for experiments that involve using different angles in infrared spectroscopy. Furthermore, the experimental methodology allowed us to derive the {\it effective} refraction index values in the infrared range for each ice component. Existing corrections typically consider geometric configurations but overlook the refractive index of the ice ($n$), a factor dependent on ice composition. The study reveals that the incident angle and the refractive index, determine the pathlength of the IR beam across the ice sample. This insight challenges conventional corrections, impacting the integrated absorption values of the IR bands and column densities. In addition, for certain ice components, variations in the incidence angle affect the longitudinal (LO) and transverse (TO) optical modes of the ice, leading to observable changes in the IR band profiles that provide information on the amorphous or crystalline structure of the ice. The practical implications of this work apply to experimental setups where normal IR measurements are unfeasible. Researchers using, for example, the standard 45$^{\circ}$ angle for IR spectroscopy, will benefit from a more accurate estimation of ice column density.
\end{abstract}


\section{Introduction} 
\label{sect:intro}
In the astrochemical literature, numerous experimental works were devoted to ice infrared (IR) spectroscopy for in situ characterization of the ice structure and monitoring of ice processes during irradiation with UV or X-rays, ion bombardment or controlled warm-up. A clear advantage of IR spectroscopy is the comparison to observed spectra of ice in solar system objects in reflectance, or ice  mantles in the coldest regions of interstellar clouds and protoplanetary disks observed in absorbance. In the later observations, the column density of the more abundant ice components along the line of sight can be determined. This revealed an ice mantle composition dominated by water and other simple molecules that include CO, CO$_2$, CH$_3$OH, OCN$^-$, OCS, H$_2$CO, HCOOH, CH$_4$, and NH$_3$, or NH4$^+$, e.g. \cite{Obert2011}.\\

The present paper aims to study the infrared (IR) spectroscopy of common molecular components in interstellar and circumstellar ice mantles, these are CO, CO$_2$, H$_2$O, CH$_3$OH, NH$_3$, CH$_4$, and H$_2$S at different angles of incidence of the IR beam with respect to the substrate used for ice deposition in our experiments. This incident angle determines the pathlength of the ice intersected by the IR beam, which corresponds to integrated absorption values and column densities larger than the ones measured at normal incidence angle. An application of this work is therefore the correct estimation of the column density in experiments using different incident angles for IR spectroscopy. The objective is to provide a value of the column density that corresponds to the one obtained at normal incident angle, because such value is directly related to the thickness of the ice if porosity is discarded. Typically, the corrections reported in the literature only account for the trivial geometrical configuration of the system, i.e. $N_0 = N_{\alpha} cos({\alpha})$ where $N_0$ is the ice column density measured at normal incidence of the IR beam and $N_{\alpha}$ the one measured at incident angle $\alpha$. These works do not take the refractive index of the ice, $n$, into consideration, which depends mainly on the ice composition, and together with the incident angle, determines the pathlength of the IR beam across the ice sample. The IR absorbance of a thin sample measured at normal incidence with a linearly polarized beam has been studied in detail by \cite{itoh2009}.\\

Transmission IR spectroscopy of thin ice films, performed at incident angle normal to the substrate, allows to observe the vibrations taking place in the plane of the substrate or with vector components in that plane. For other incident angles, the IR beam may interfere with both the surface-parallel and surface-perpendicular vibrational modes. Therefore, even for unpolarized IR beam spectroscopy, which is the technique most commonly used in ice studies, the incident angle can influence the value of the integrated absorption, and also the band profiles to some degree, since these parameters depend on the vibrations activated with IR light. For instance, the so-called longitudinal (LO) and transverse (TO) optical modes of CO ice fall near 2142 and 2138 cm$^{-1}$ (e.g., \cite{Palumbo2006}). Unpolarized IR light at normal incidence has electric field components parallel to the film surface, and is therefore only sensitive to the TO mode. On the other hand, at oblique angles both the LO and TO modes can be observed. Such LO-TO splitting is attributed to dipole-dipole interactions in CO ice \citep{Zumofen1978, Lasne2015}; it is most evident in crystalline ice samples but may also be seen in at least partially disordered ice (\cite{MunozCaro2016}, and ref. therein).\\

As discussed above, IR spectroscopy of ices at oblique incidence is used to infer more information from the ice structure, or simply because either the configuration of the vacuum chamber or the experimental protocol do not allow IR spectroscopy at normal incidence. This work presents a methodical study of the effects of incident angle on ice spectroscopy with unpolarized IR light. For this, different incident angles of the IR beam with respect to the ice substrate were explored in the reported experiments.\\

\section{Experimental} 
\subsection{Experimental setup} \label{sect:exp}
Experiments were conducted using the Interstellar Astrochemistry Chamber (ISAC), which is an ultra-high vacuum (UHV) chamber designed to simulate the conditions present in the interstellar medium (ISM), such as temperature, low particle density, and ultraviolet (UV) radiation field. The base pressure of ISAC is 4x10$^{-11}$ mbar. Further details about ISAC can be found in \cite{MunozCaro2010}.\\

The cold finger of the chamber is cooled down using a closed-cycle He cryostat to a temperature close to 8 K, where a sample holder with a MgF$_2$ window acts as the substrate for ice deposition. However, in the experiments reported in this work, the lowest achievable temperature was 11 K because the radiation shield surrounding the sample holder was removed during these experiments, allowing IR measurements at a wider range of angles. The temperature of the ice sample was measured by placing a silicon diode sensor just below the ice sample and attaching it to the sample holder. The temperature is controlled by a Lakeshore 331 temperature controller with a precision of 0.1 K. For these experiments, the temperature remained constant at 11 K. The gas line system in ISAC allows the introduction of gas species with a controlled composition, monitored by Quadrupole Mass Spectrometry (QMS, Pfeiffer Vacuum, Prisma QMS 200). In the main chamber of ISAC, pressure was monitored using a Bayard-Alpert gauge located approximately 23 cm below the deposition plane. The experimental parameters were recorded during the experiments to ensure constant conditions.\\

Fig. \ref{fig.ISAC_scheme} shows a scheme of the ISAC set-up. During deposition, the gas valve is opened and the gas is directed to the cold substrate via a deposition tube. The end of this tube is about 3 cm from the substrate, and the sample holder can be rotated at any desired angle using a motorised system. The following abundant molecular components of interstellar and circumstellar ice mantles were introduced into the main chamber for 10 minutes: CO, CO$_2$, H$_2$O, CH$_3$OH, NH$_3$, CH$_4$, and H$_2$S (one pure component in each experiment). This was achieved by opening the gas line 2 leak valve, see Fig. \ref{fig.ISAC_scheme}, to attain the desired deposition pressure of 2 $\times$ 10$^{-7}$ mbar.\\

In these experiments, the angle of deposition with respect to the cold substrate was consistently set at $\theta_i$ = 0$^{\circ}$. This ensured a perpendicular incidence of the deposited molecules onto the substrate surface, allowing for uniform and controlled ice deposition \citep{GonzalezDiaz2019}.
After the completion of the deposition of the aforementioned molecular ice components, infrared spectra were recorded using Fourier-transform infrared spectroscopy (FTIR) in transmittance mode, employing a Bruker Vertex 70 spectrometer operating at a working spectral resolution of 2 cm$^{-1}$. These spectra were collected at different angles of incidence from 0 to 60$^{\circ}$, the latter is the maximum oblique angle allowed in ISAC for IR spectroscopy.\\

\subsection{Column density estimation versus incident angle of the IR beam}
\label{sec:Column density estimations versus incident angles of the IR beam}
\begin{figure}
    \centering
    \includegraphics[width=\textwidth]{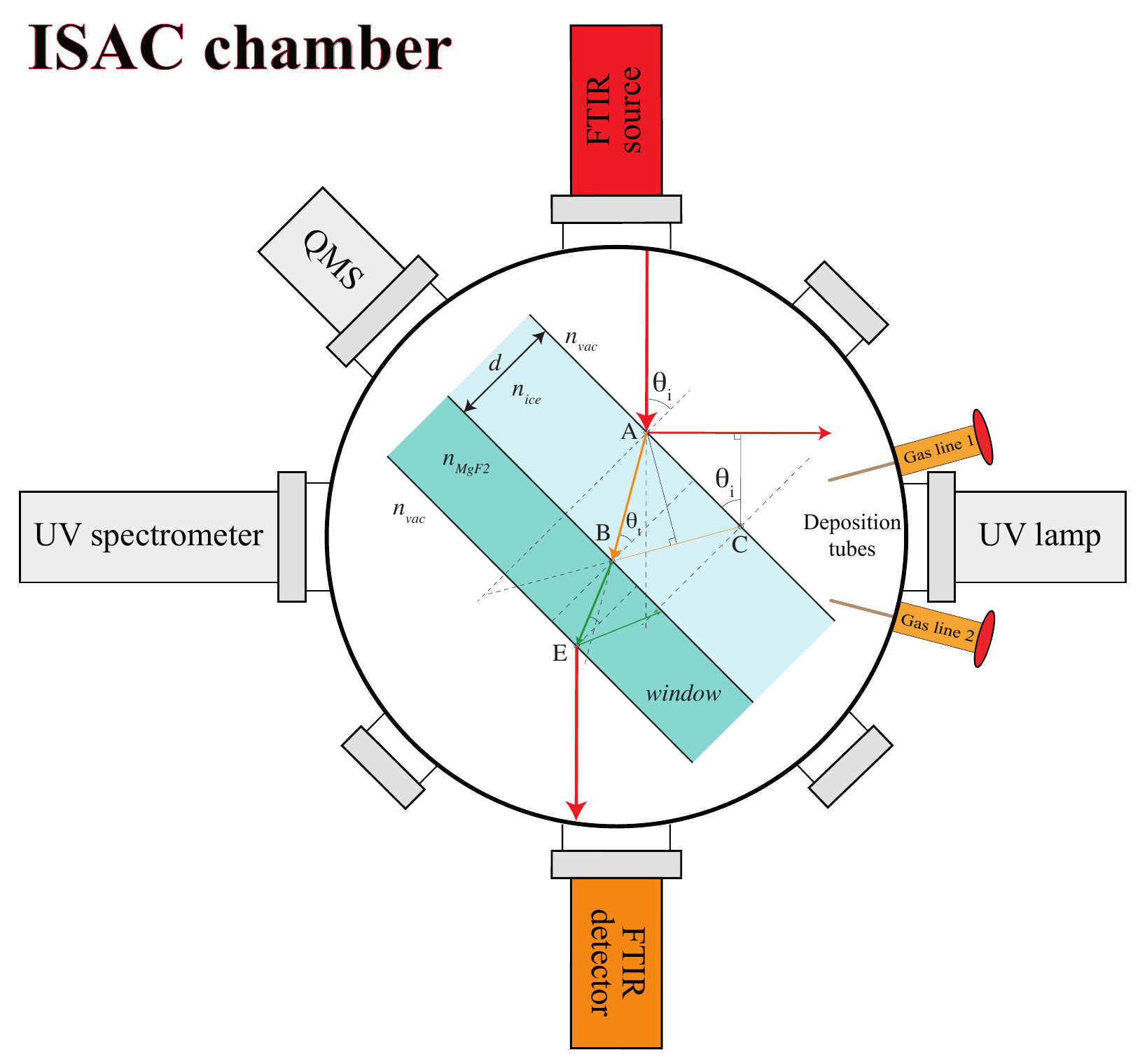}
     \caption{Sketch of the path followed by the IR beam as a function of the incident angle. The positions of the detectors are shown. QMS stands for quadrupole mass spectrometer, FTIR is the Fourier transform infrared spectrometer. The FTIR spectrometer with the IR source is facing one viewport and the FTIR detector is on the opposite side. The UV spectrometer is located at the opposite side of the ice sample.}
    \label{fig.ISAC_scheme}
\end{figure}

Figure \ref{fig.ISAC_scheme} represents the path of the IR beam that passes through the deposited ice, red trace, starting from point $A$ and ending at point $E$. The substrate used for ice deposition is a MgF$_2$ window, which has minimal absorption in the mid-IR range displayed in the reported experiments. The line segment $|AB|$ corresponds therefore to the distance travelled by the IR beam through the ice layer. This distance is crucial in determining the absorption and the column density of the IR beam, depending on the incident angle.\\

From the triangle formed by $ABC$ it follows that
\begin{equation}
    |AB| = |BC| = \frac{d}{\cos{\theta_t}}
\end{equation}

Using Snell's law,
\begin{equation}
    n_{vac} \sin{\theta_i} = n_{ice} \sin{\theta_t}
\end{equation}

we obtain 

\begin{align}
    |AB| &= \frac{d}{cos(arcsin(\frac{n_{vac}sin{\theta_i}}{n_{ice}})} \label{ABpath}
\end{align}

If the medium surrounding the ice layer is air or vacuum, $n_{vac}$ = 1, 
equation \ref{ABpath} can be rearranged using 
$\cos{(\arcsin{x})} = \sqrt{1-x^2}$ to obtain

\begin{align}
    |AB| &= \frac{d\cdot n_{ice}}{\sqrt{n_{ice}^2-sin^2{\theta_i}}} \label{ABpathSimpli}
\end{align}

Equation \ref{ABpathSimpli} can be expressed in its linear form $y = a\cdot x$ 

\begin{align}
     sin{\theta_i} &= n_{ice} \cdot \sqrt{1-\frac{d^2}{|AB|^2}} \label{AB_sin(theta)}
 \end{align}

where $y = sin{\theta_i}$, $x=\sqrt{1-\frac{d^2}{|AB|^2}}$, and $a = n_{ice}$ is the index of refraction of the ice that corresponds to the slope of the linear fit. If the data recorded by the FT-IR spectrometer is normalised, then $d = 1$.\\

Finally, in IR spectra acquired at incident angle $\theta_i$, the correct value of the ice column density $N_0$ as if it were measured at normal incidence, can be easily obtained from the above equations:\\

\begin{align}
     N_0 = N_{\theta_i} \frac{|AB|_0}{|AB|_{\theta_i}}= N_{\theta_i} { \sqrt{1 - \frac{sin^2{\theta_i}}{n_{ice}^2}}  }
     \label{N_corrected}
 \end{align}

%
%

Hence, Equation \ref{AB_sin(theta)} can be used to determine the absorption and column density at different incident angles. Additionally, from the slope of the linear fit, an {\it effective} refraction index, $n_{ice}$, can be derived.\\

\section{Results} \

The spectroscopy performed at different incident angles of the more abundant ice components in interstellar ice mantles is presented as follows. First, the results are shown for the CO and CO$_2$ ices. These ice samples exhibit clear effects related to the longitudinal (LO) and transverse (TO) optical modes when IR spectra are acquired at oblique incident angles. Second, H$_2$O, CH$_3$OH, NH$_3$, CH$_4$ and H$_2$S spectroscopy is presented. In general, these ice spectra only express the increase of the incident IR angle as an enhancement of the IR absorption bands. The growing IR absorption of these ices is simply explained by the larger IR beam path across the ice at oblique angles compared to normal incidence.\\          

\subsection{CO} 
\label{CO}
\begin{figure}
    \centering
    \includegraphics[width=\textwidth]{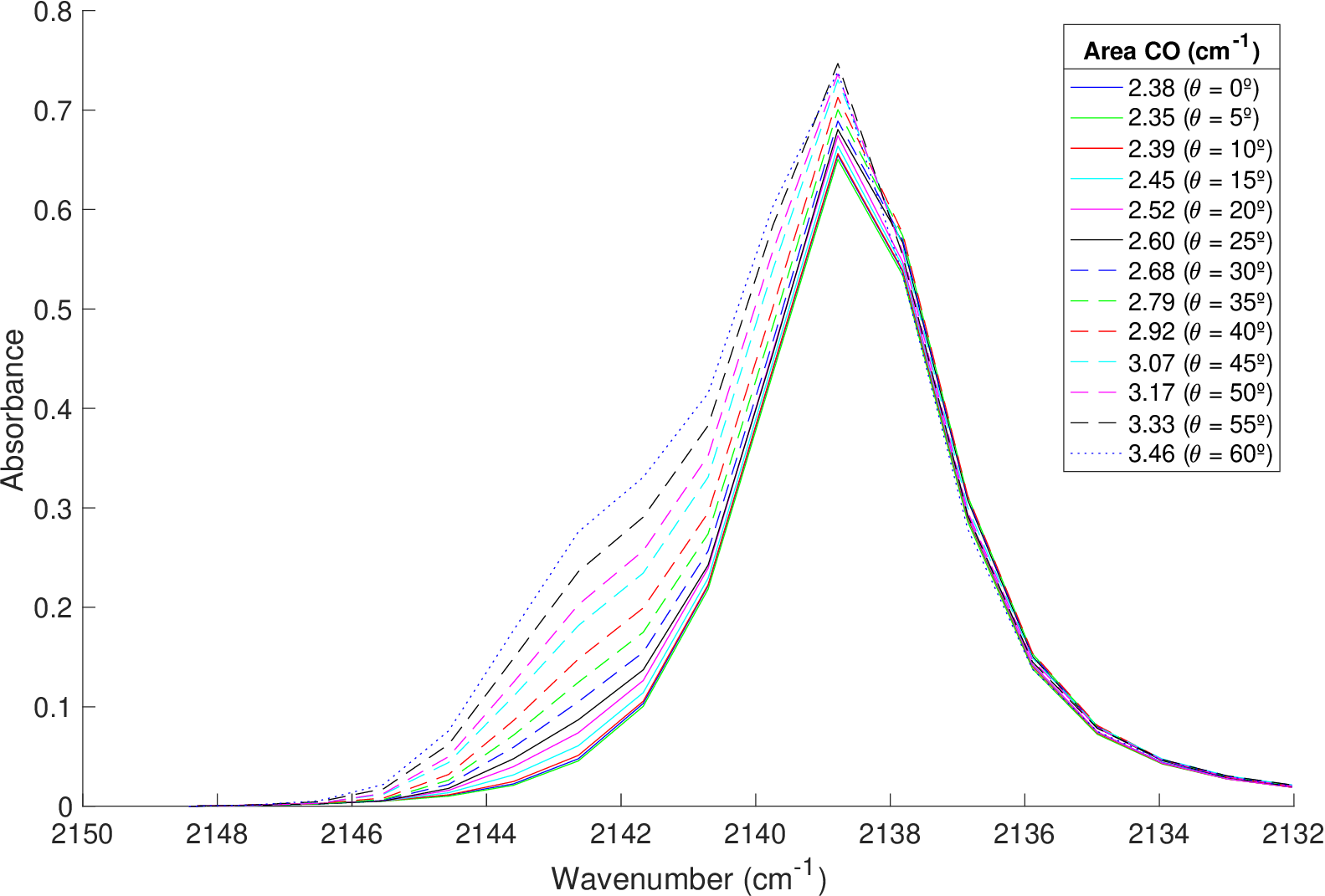}
       \caption{IR spectra of CO ice deposited at 11 K as a function of the incidence angle used for IR spectrometry.}
    \label{fig:IR_CO_Trap}
\end{figure}

As was mentioned in Sect.\ref{sect:intro}, the TO mode around 2138 cm$^{-1}$ is the dominant one; see Fig. \ref{fig:IR_CO_Trap}, while the LO mode around 2142 cm$^{-1}$ gradually becomes observable as the incident angle increases up to about 40$^{\circ}$, resembling a "wing" of the TO mode at even larger angles. The appearance of the LO mode follows a slight blueshift of the IR band.\\

\begin{figure}
    \centering
    \includegraphics[width=\textwidth]{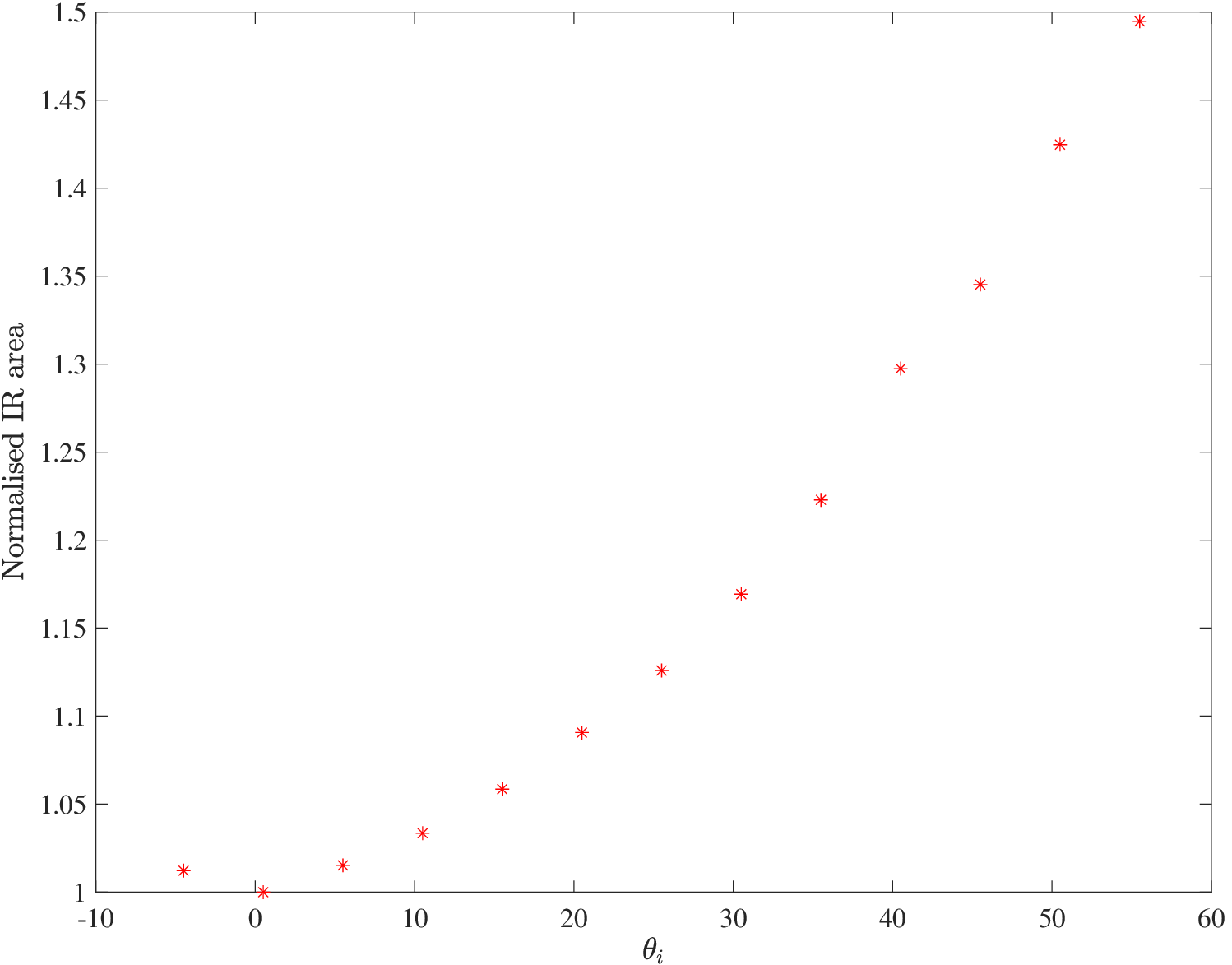}
    \caption{Normalised  values of the integrated IR absorption spectra, corresponding to the IR band areas of CO ice deposited at 11 K as a function of the incident angle $\theta_i$.}
    \label{fig:IR_CO_non_Linear}
\end{figure}

Fig. \ref{fig:IR_CO_non_Linear} illustrates the normalised values of the integrated IR absorption spectra of CO ice deposited at 11 K, plotted against the incident angle $\theta_i$. Notably, as the incident angle increases, there is a corresponding increase in the normalised value of the integrated IR absorption. This phenomenon can be attributed to the increase of the IR path traversing through the deposited ice layer for oblique incident angles.\\

\begin{figure}
    \centering
    \includegraphics[width=\textwidth]{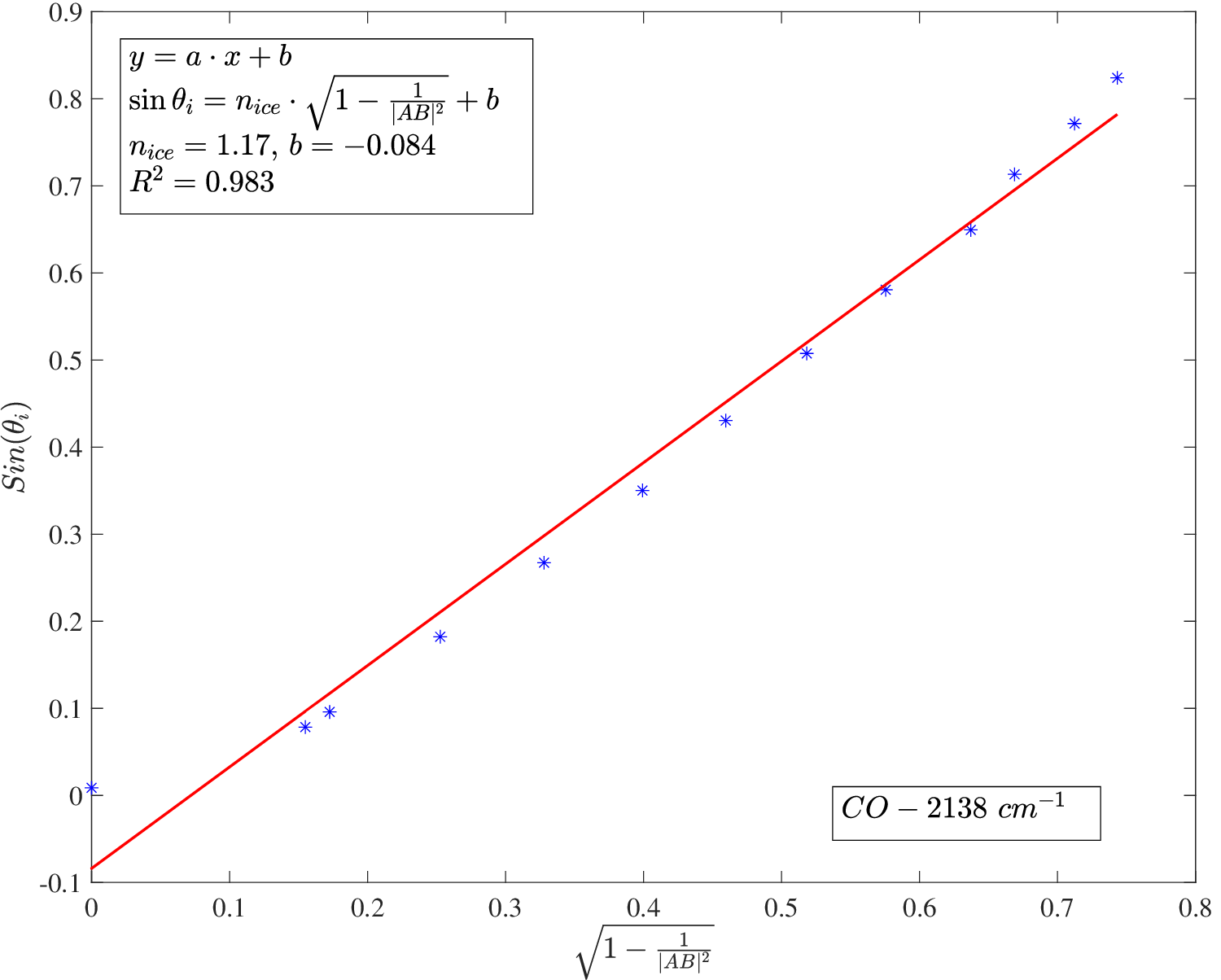}
        \caption{Representation of sine of incident angle of the IR beam, $sin{\theta_i}$, for CO ice deposited at 11 K as a function of $\sqrt{1-\frac{d^2}{|AB|^2}}$. The red trace is the linear fit.}
    \label{fig:IR_CO_Linear}
\end{figure}

Fig. \ref{fig:IR_CO_Linear} shows $sin{\theta_i}$ as a function of the $\sqrt{1-\frac{d^2}{|AB|^2}}$ values with a linear fit. This fit provides a value of $n_{ice}$ = 1.17 which should be regarded as the {\it effective} index of refraction for CO ice deposited at 11 K in these experiments. This value of $n_{ice}$ applies to the 2138 cm$^{-1}$ band of CO and probably corresponds to an average value for the integrated absorption range of this band. The same holds for the other ice compositions reported here below. These $n_{ice}$ values are provided in Table \ref{tab:table_Resumen} and will be discussed in Sect. \ref{sect:conc}.    

\subsection{CO$_2$} 
\label{CO2}
\begin{figure}
    \centering
    \includegraphics[width=\textwidth]{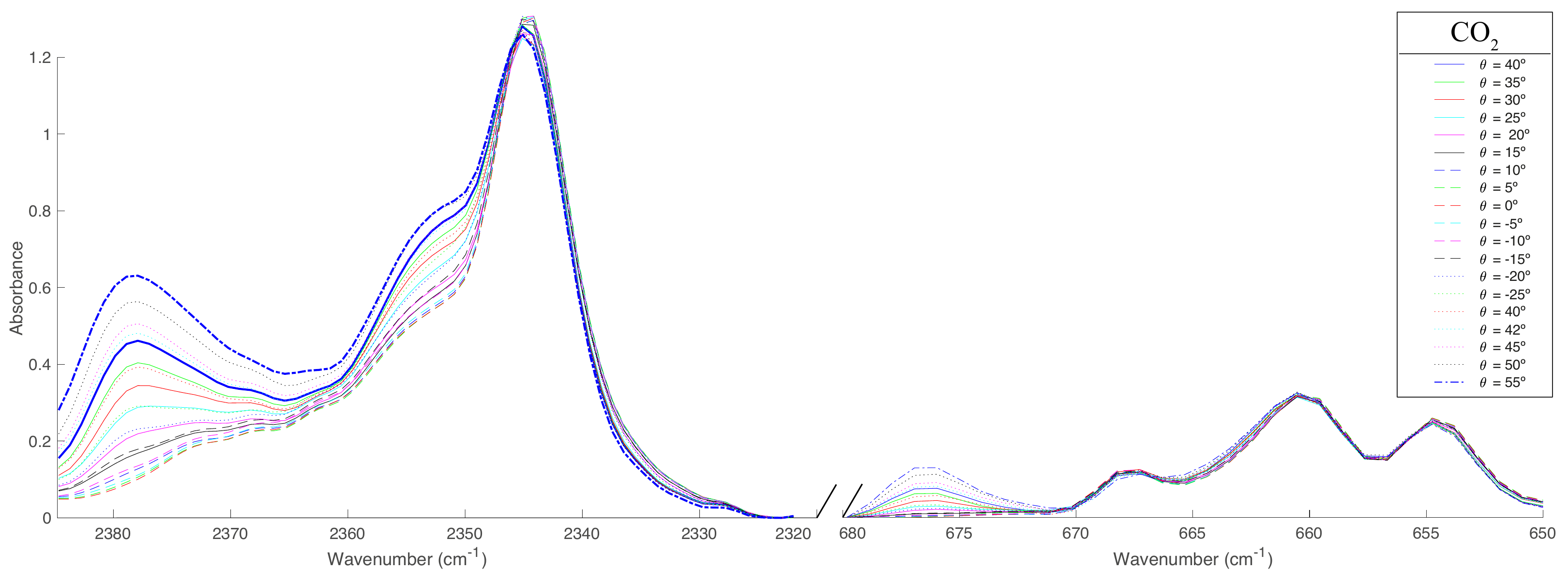}
    \caption{Infrared spectra of CO$_2$ ice deposited at 11 K measured at  different incident angles.}
    \label{fig:IR_CO2}
\end{figure}

Fig. \ref{fig:IR_CO2} shows the IR spectra of pure CO$_2$ ice at different incident angles. The antisymmetric stretching fundamental $\nu_3$ mode near 2343 cm$^{-1}$ is little incident angle dependent, while the bands near 2351 cm$^{-1}$ and 2380 cm$^{-1}$ show an increasing absorption with the incident angle. The 2351 cm$^{-1}$ band is related to the degree of porosity in the ice \citep{Schiltz_2024}.

\begin{figure}
    \centering
    \includegraphics[width=\textwidth]{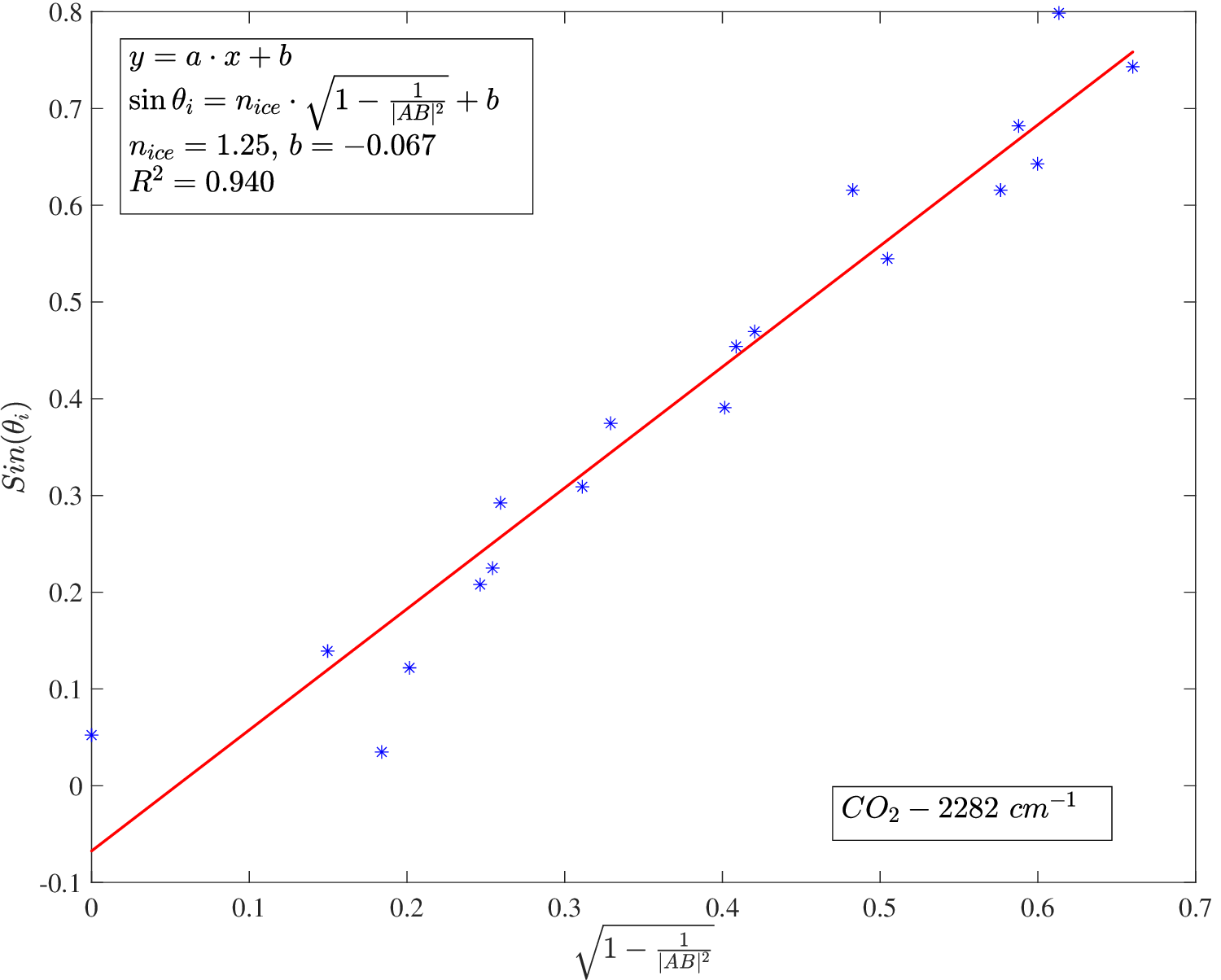}
    \caption{Representation of sine of incident angle of the IR beam, $sin{\theta_i}$, for CO$_2$ ice deposited at 11 K as a function of $\sqrt{1-\frac{d^2}{|AB|^2}}$. Data corresponds to the band around 2282 cm$^{-1}$. The red trace is the linear fit.}
    \label{fig:IR_CO2_2_Gauss1_Absorbance}
\end{figure}

\begin{figure}
    \centering
    \includegraphics[width=\textwidth]{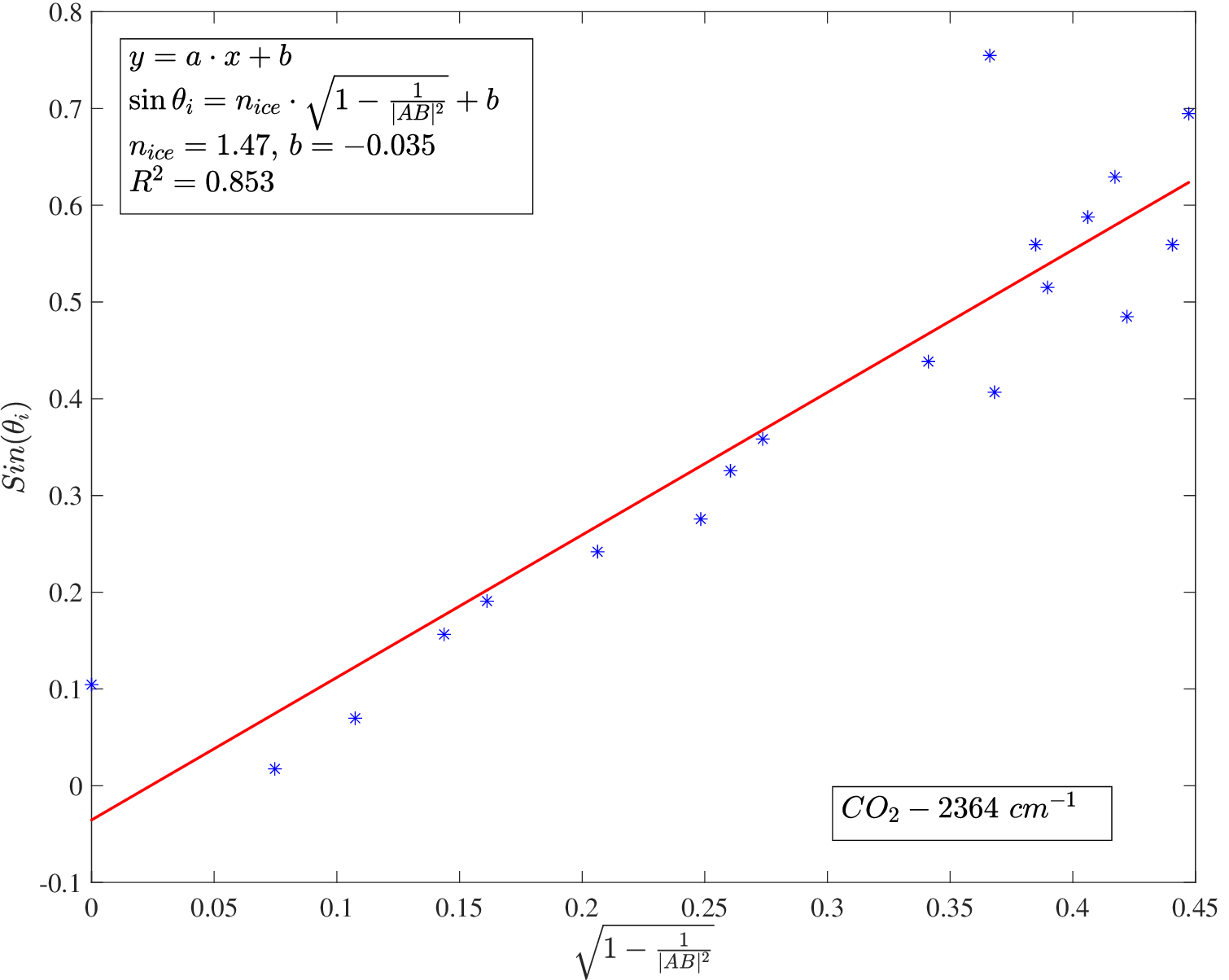}
    \caption{Representation of sine of incident angle of the IR beam, $sin{\theta_i}$, for CO$_2$ ice deposited at 11 K as a function of $\sqrt{1-\frac{d^2}{|AB|^2}}$. Data corresponds to the band around 2364 cm$^{-1}$. The red trace is the linear fit.}
    \label{fig:IR_CO2_linear}
\end{figure}

\begin{figure}
    \centering
    \includegraphics[width=\textwidth]{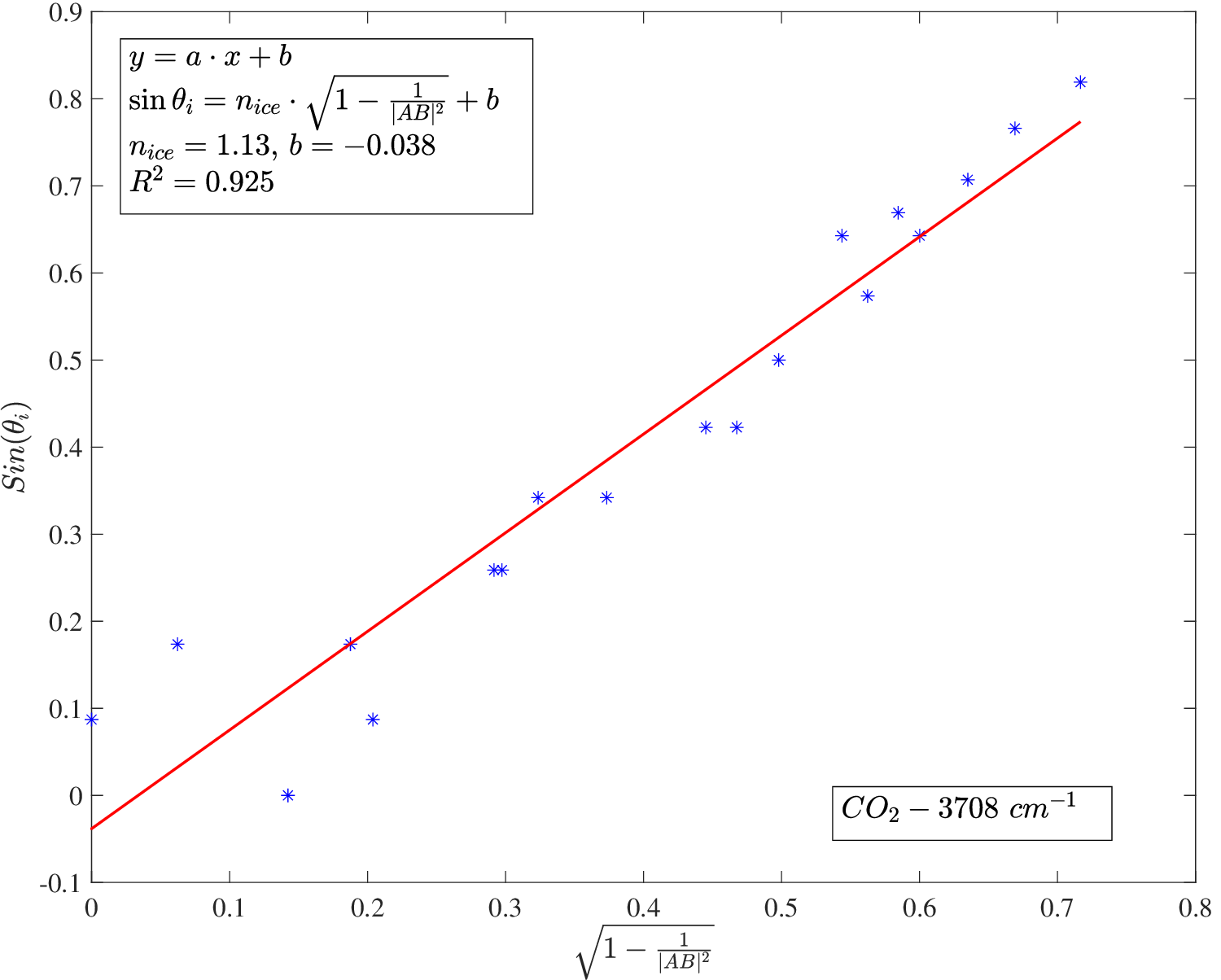}
    \caption{Representation of sine of incident angle of the IR beam, $sin{\theta_i}$, for CO$_2$ ice deposited at 11 K as a function of $\sqrt{1-\frac{d^2}{|AB|^2}}$. Data corresponds to the band around 3708 cm$^{-1}$. The red trace is the linear fit.}
    \label{fig:IR_CO2_3_Gauss1_Absorbance}
\end{figure}

The 2380 cm$^{-1}$ band corresponds to the longitudinal optical (LO) mode. It is often unnoticed in amorphous CO$_2$ ice spectra measured at normal incidence with an unpolarized IR beam (e.g., \cite{Falk1987}), and it is best observed in the crystalline ice form. However, this feature was observed using
reflection-absorption IR (RAIR) spectroscopy for amorphous CO$_2$ deposited at 14 K with radiation polarised in the plane of propagation of the incident IR beam (P-polarised), indicating that the vibration of the 2380 cm$^{-1}$ band is in the normal direction to the substrate or has vector components along that direction, as expected for the LO mode (\cite{Escribano2013}). This orientation of the 2380 cm$^{-1}$ band explains its increasing intensity for more oblique angles in Fig. \ref{fig:IR_CO2}. \cite{Escribano2013} explained that the CO$_2$ ice sample  displays some crystalline character due to the layer
accumulation (ice thickness increase) even at this low temperature, as evidenced by the splitting of
the $\nu_2$ (bending) modes. This argumentation is also valid for our experiments. Because the band around 2351 cm$^{-1}$ follows the same trend as the 2380 cm$^{-1}$ band, its corresponding vibration has a component in the direction normal to the ice surface. This preferred orientation, in addition to the association of this band with porosity in CO$_2$ ice, indicates that the plausible vibration is a dangling mode in CO$_2$ molecules \citep{Schiltz_2024}. While danglings at the surface of the ice will follow the normal direction, danglings of molecules on pore walls may also be oriented differently with respect to the cold substrate. These danglings would account for the smaller absorption of the 2351 cm$^{-1}$ band measured at normal incidence. As for the CO case, Fig. \ref{fig:IR_CO2_linear} shows the linear fit of the amorphous CO$_2$ ice spectral data that allows estimation of the \textit{effective} index of refraction, $n_{CO_2}$ = 1.47, for the 2364 cm$^{-1}$ absorption. Fig. \ref{fig:IR_CO2_2_Gauss1_Absorbance} and Fig. \ref{fig:IR_CO2_3_Gauss1_Absorbance} show the corresponding linear fits for the 2282 and 3708 cm$^{-1}$ IR bands, respectively. The goodness of fit is acceptable, $R$ = 0.940, 0.853 and 0.925, for the three bands.\\ 

\subsection{H$_2$O} \label{H2O}

\begin{figure}
    \centering
    \includegraphics[width =\textwidth]{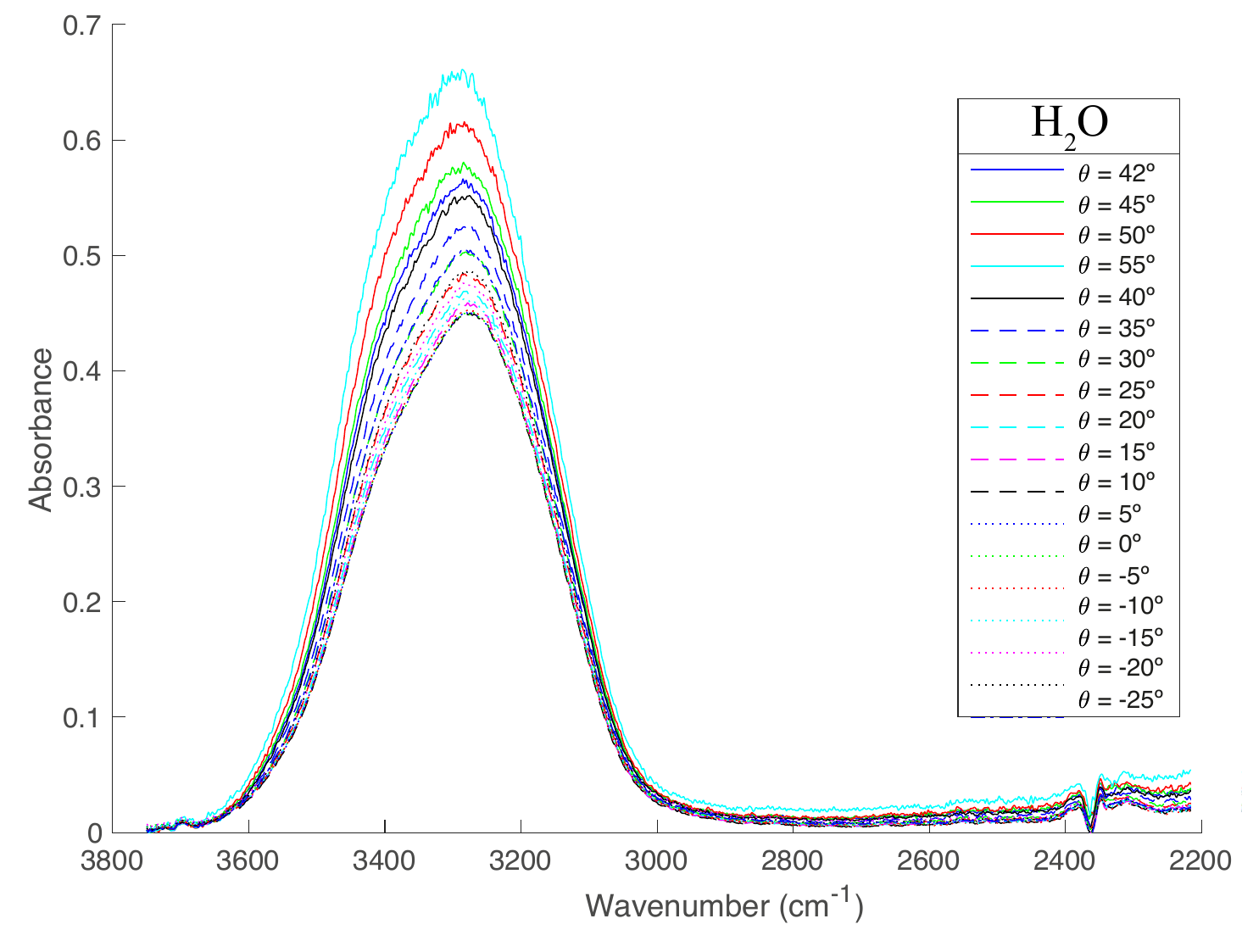}
    \caption{IR band around 3283 cm$^{-1}$ of H$_2$O ice deposited at 11 K for various angles of incidence.}
    \label{fig:trap_h20}
\end{figure}

Fig. \ref{fig:trap_h20} shows the IR spectra of pure H$_2$O ice at different incident angles. It is observed that the increase in the angle of incidence is accompanied by an increasing absorbance. As already mentioned, this is due to the larger IR beam path going across the ice layer for increasing incident angles. The maximum absorbance occurs at the most oblique angle that could be measured in this experiment, 55$^\circ$.

\begin{figure}
    \centering
    \includegraphics[width =\textwidth]{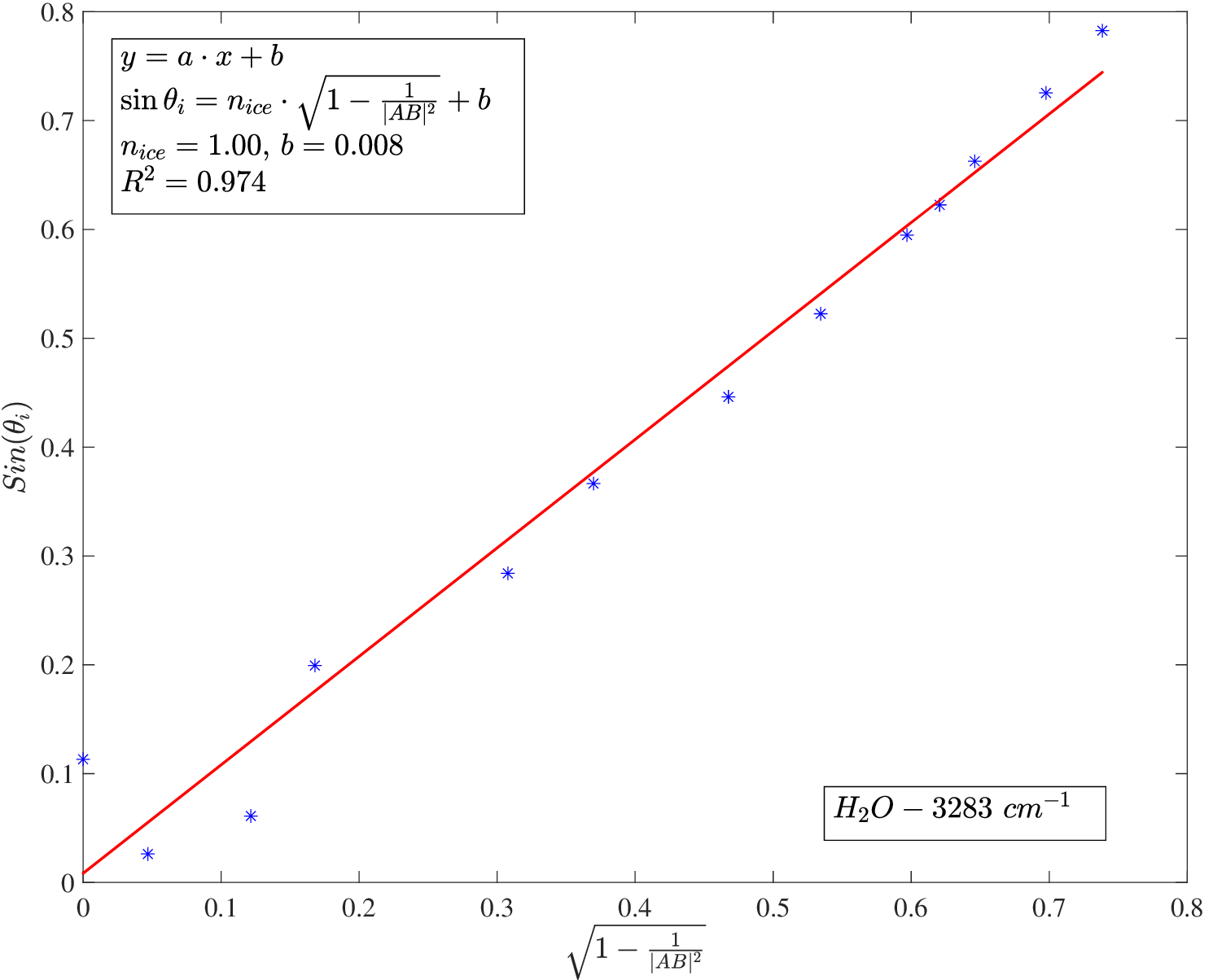}
    \caption{Representation of sine of incident angle of the IR beam, $sin{\theta_i}$, for H$_2$O ice deposited at 11 K as a function of $\sqrt{1-\frac{d^2}{|AB|^2}}$. The red trace is the linear fit.}
    \label{fig:LinearH2O_H20}
\end{figure}

Fig. \ref{fig:LinearH2O_H20} shows $sin{\theta_i}$ as a function of the $\sqrt{1-\frac{d^2}{|AB|^2}}$ in H$_2$O ice spectra and the linear data fit. This data allows the estimation of the index of refraction, $n_{H_2O}$ = 1.00. The goodness of fit is acceptable, $R$ = 0.974. It should be regarded as the {\it effective} index of refraction for H$_2$O ice deposited at 11 K in these experiments. This value applies to the 3283 cm$^{-1}$ band of H$_2$O ice. 

\subsection{CH$_3$OH} 
\label{CH3OH}

\begin{figure}
    \centering
    \includegraphics[width=\textwidth]{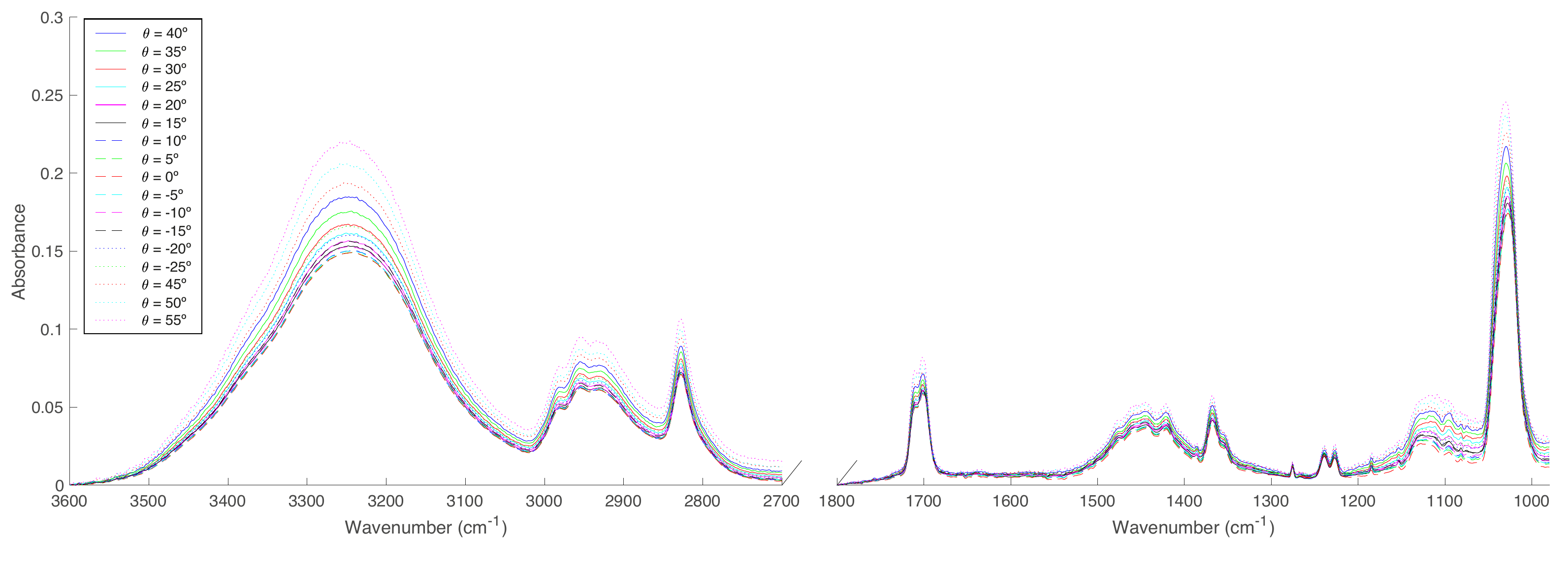}
    \caption{IR spectra of CH$_3$OH ice deposited at 11 K as a function of the incident angle used for IR spectrometry.}
    \label{fig:trap_CH3OH_allRange}
\end{figure}

\begin{figure}
    \centering
    \includegraphics[width =\textwidth]{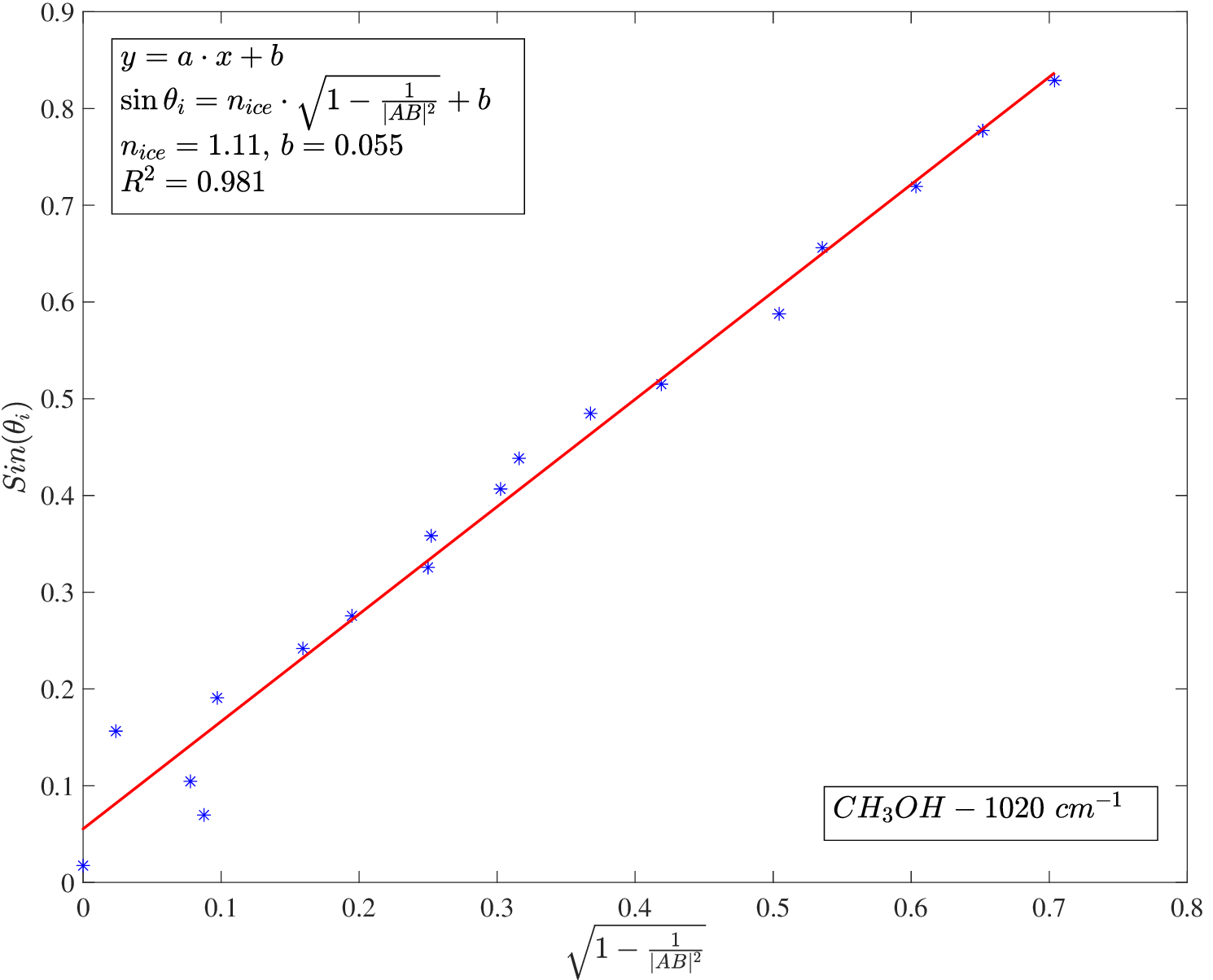}
    \caption{Representation of sine of incident angle of the IR beam, $sin{\theta_i}$, for CH$_3$OH ice deposited at 11 K as a function of $\sqrt{1-\frac{d^2}{|AB|^2}}$. Data corresponds to the band around 1020 cm$^{-1}$. The red trace is the linear fit.}
    \label{fig:LinearCH3OH_1020_CH3OH_1020}
\end{figure}

\begin{figure}
    \centering
    \includegraphics[width =\textwidth]{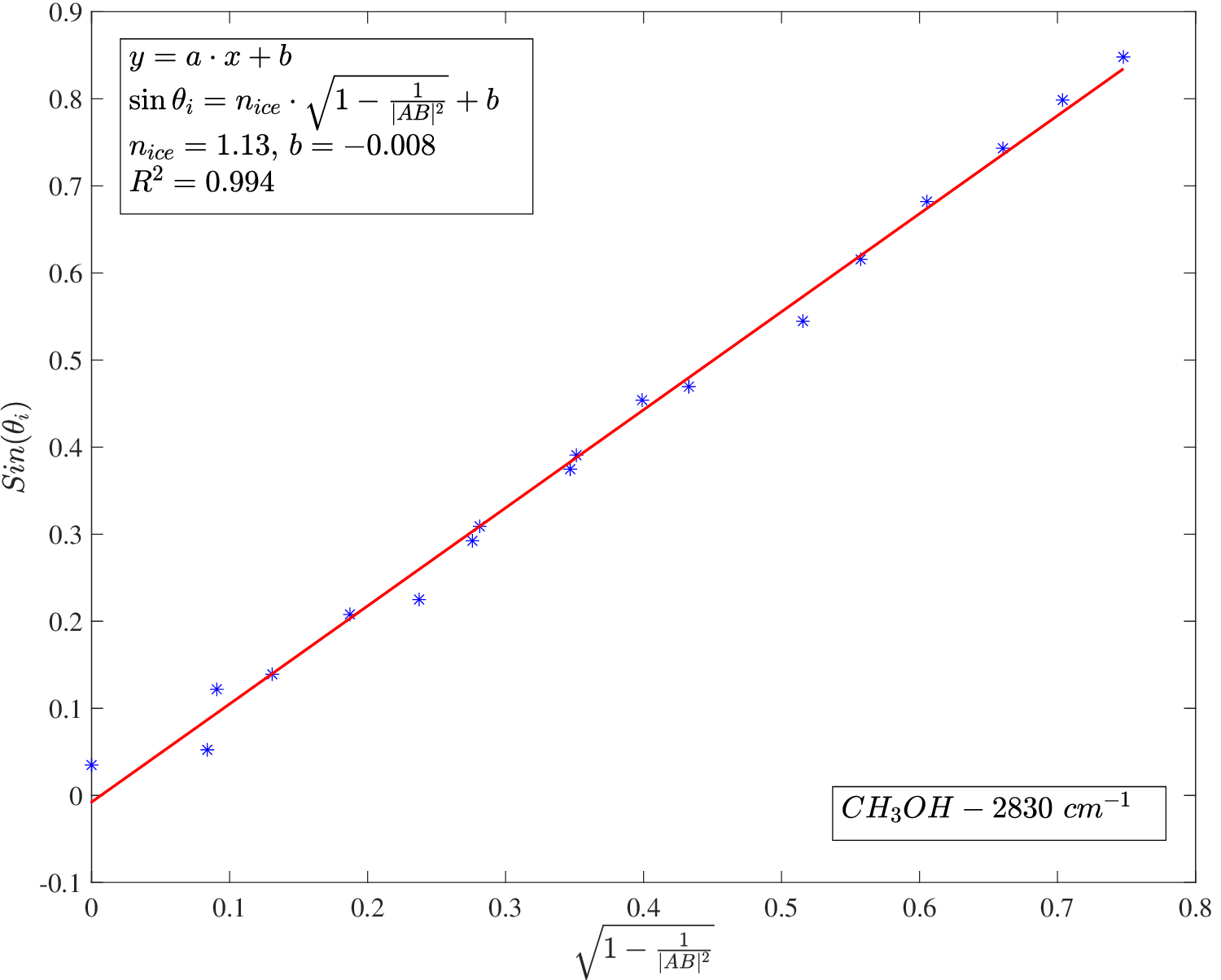}
    \caption{Representation of sine of incident angle of the IR beam, $sin{\theta_i}$, for CH$_3$OH ice deposited at 11 K as a function of $\sqrt{1-\frac{d^2}{|AB|^2}}$. Data corresponds to the band around 2830 cm$^{-1}$. The red trace is the linear fit.}
    \label{fig:LinearCH3OH_2830_CH3OH_2830}
\end{figure}

\begin{figure}
    \centering
    \includegraphics[width =\textwidth]{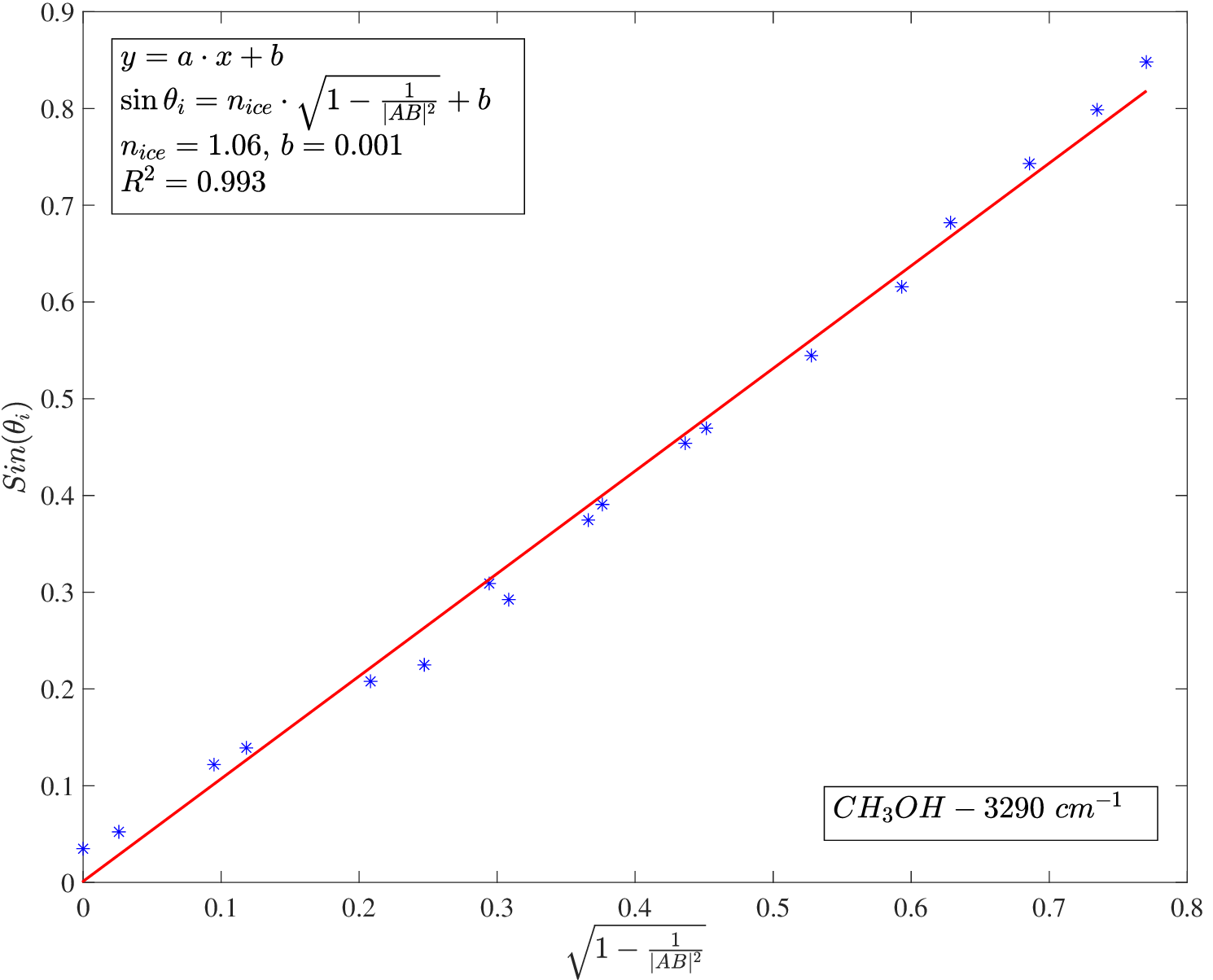}
    \caption{Representation of sine of incident angle of the IR beam, $sin{\theta_i}$, for CH$_3$OH ice deposited at 11 K as a function of $\sqrt{1-\frac{d^2}{|AB|^2}}$. Data corresponds to the band around 3290 cm$^{-1}$. The red trace is the linear fit.}
    \label{fig:LinearCH3OH_3290_CH3OH_3290}
\end{figure}

Fig. \ref{fig:trap_CH3OH_allRange} shows the IR spectra of CH$_3$OH ice at different incident angles. 
Figs. 
\ref{fig:LinearCH3OH_1020_CH3OH_1020},
\ref{fig:LinearCH3OH_2830_CH3OH_2830}, and \ref{fig:LinearCH3OH_3290_CH3OH_3290} display $sin{\theta_i}$ as a function of the $\sqrt{1-\frac{d^2}{|AB|^2}}$ for CH$_3$OH ice spectral data in the 1020, 2830 and 3290 cm$^{-1}$ bands of CH$_3$OH, respectively. From these plots, the index of refraction was estimated, $n_{CH_3OH}$ = 1.11, 1.13 and 1.06 for the 1020, 2830 and 3290 cm$^{-1}$ bands of CH$_3$OH, respectively. The goodness of fit is acceptable, $R$ = 0.981, 0.994 and 0.993. These values represent the \textit{effective} index of refraction for CH$_3$OH ice deposited at 11 K in these experiments, which corresponds to an average value over the integrated absorption range of these bands.

\subsection{NH$_3$} 
\label{NH3}

\begin{figure}
    \centering
    \includegraphics[width=\textwidth]{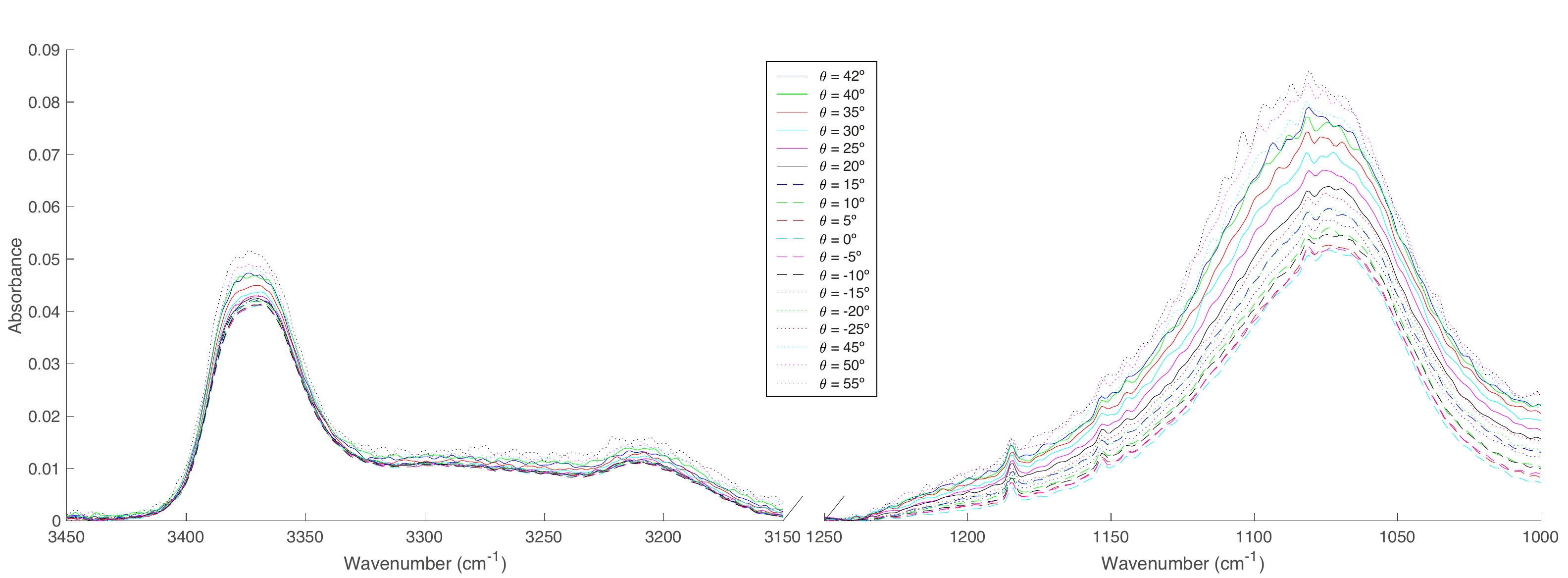}
    \caption{IR spectra of NH$_3$ ice deposited at 11 K as a function of the incident angle. The bands at 1082cm$^{-1}$ and 3375cm$^{-1}$ are displayed.}
    \label{fig:trap_NH3_All}
\end{figure}

\begin{figure}
    \centering
    \includegraphics[width=\textwidth]{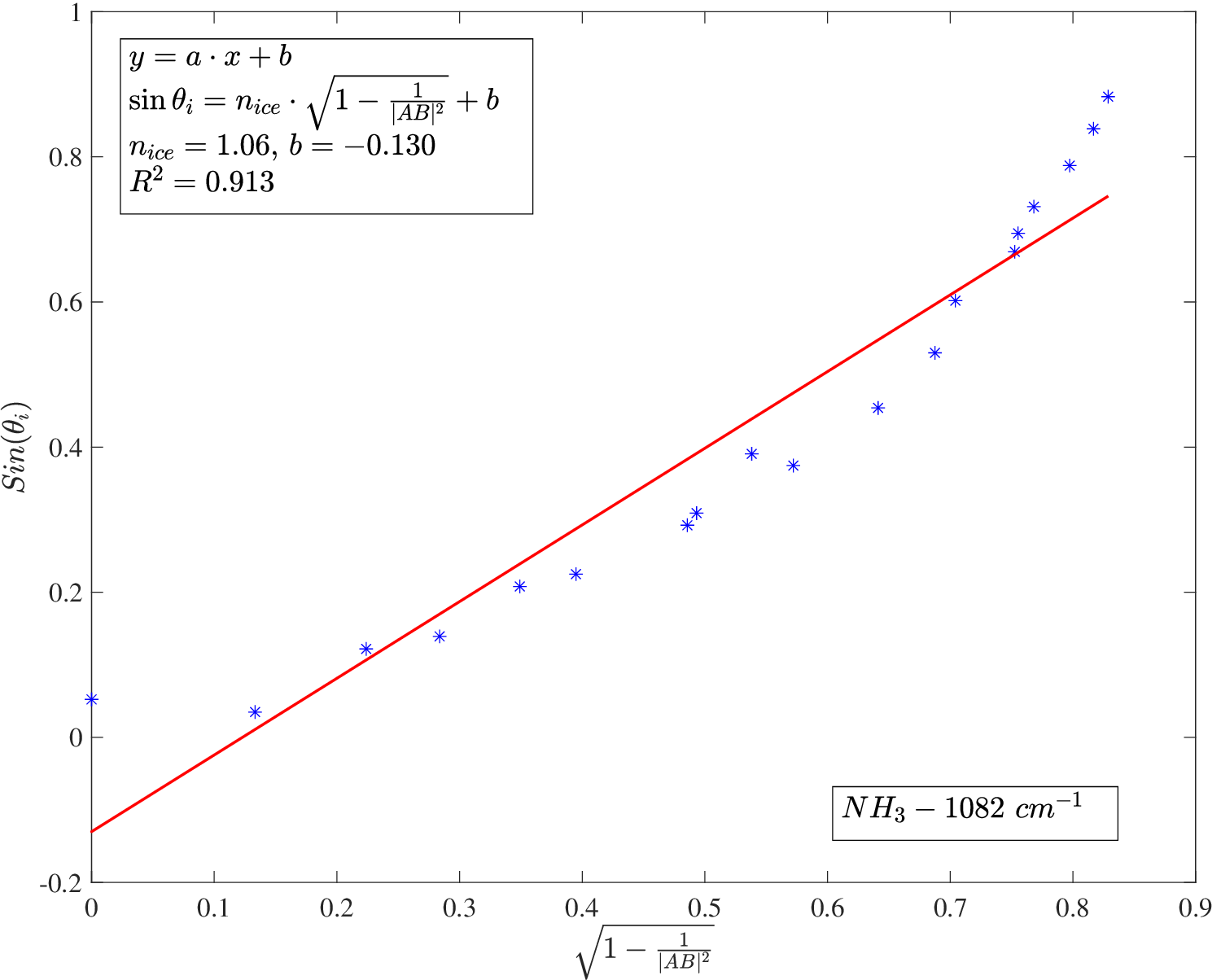}
    \caption{Representation of sine of incident angle of the IR beam, $sin{\theta_i}$, for NH$_3$ ice deposited at 11 K as a function of $\sqrt{1-\frac{d^2}{|AB|^2}}$. Data corresponds to the band around 1082 cm$^{-1}$. The red trace is the linear fit.}
    \label{fig:LinearNH3_1082}
\end{figure}

\begin{figure}
    \centering
    \includegraphics[width=\textwidth]{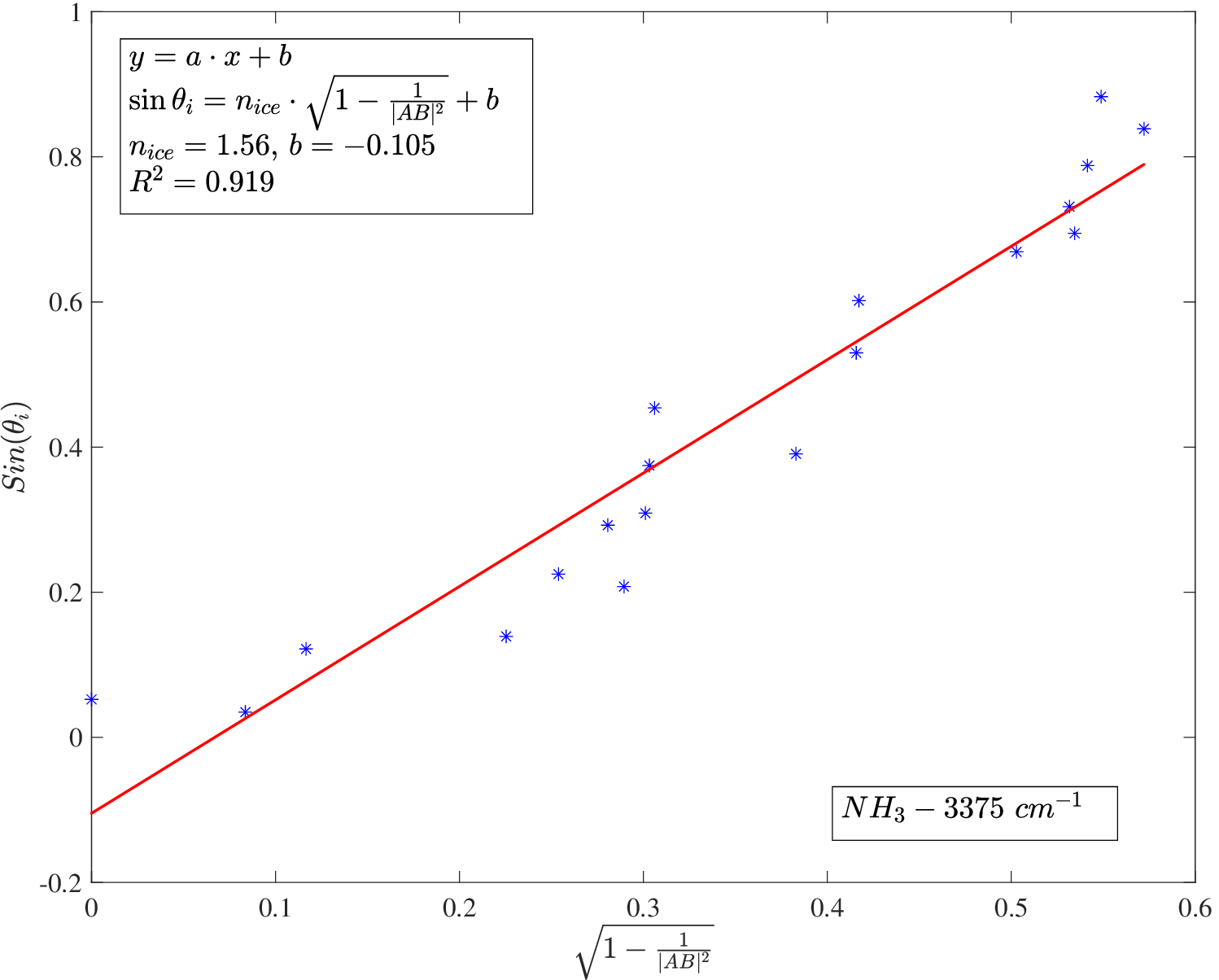}
    \caption{Representation of sine of incident angle of the IR beam, $sin{\theta_i}$, for NH$_3$ ice deposited at 11 K as a function of $\sqrt{1-\frac{d^2}{|AB|^2}}$. Data corresponds to the band around 3375 cm$^{-1}$. The red trace is the linear fit.}
    \label{fig:LinearNH3_3375}
\end{figure}

Fig. \ref{fig:trap_NH3_All} shows the IR spectra of NH$_3$ ice at different incident angles. Figs. \ref{fig:LinearNH3_1082}
and \ref{fig:LinearNH3_3375} represent $sin{\theta_i}$ as a function of the $\sqrt{1-\frac{d^2}{|AB|^2}}$ for NH$_3$ ice spectral data in the 1082 and 3375 cm$^{-1}$ bands of NH$_3$, respectively. From these plots, the index of refraction are estimated, $n_{NH_3}$ = 1.06 and 1.56 for the 1082 and 3375 cm$^{-1}$ band of NH$_3$. The goodness of fit is acceptable, $R$ = 0.913 and 0.919. These values represent the {\it effective} refractive index for NH$_3$ ice deposited at 11 K in these experiments.

\subsection{CH$_4$} 
\label{CH4}

\begin{figure}
    \centering
    \includegraphics[width=\textwidth]{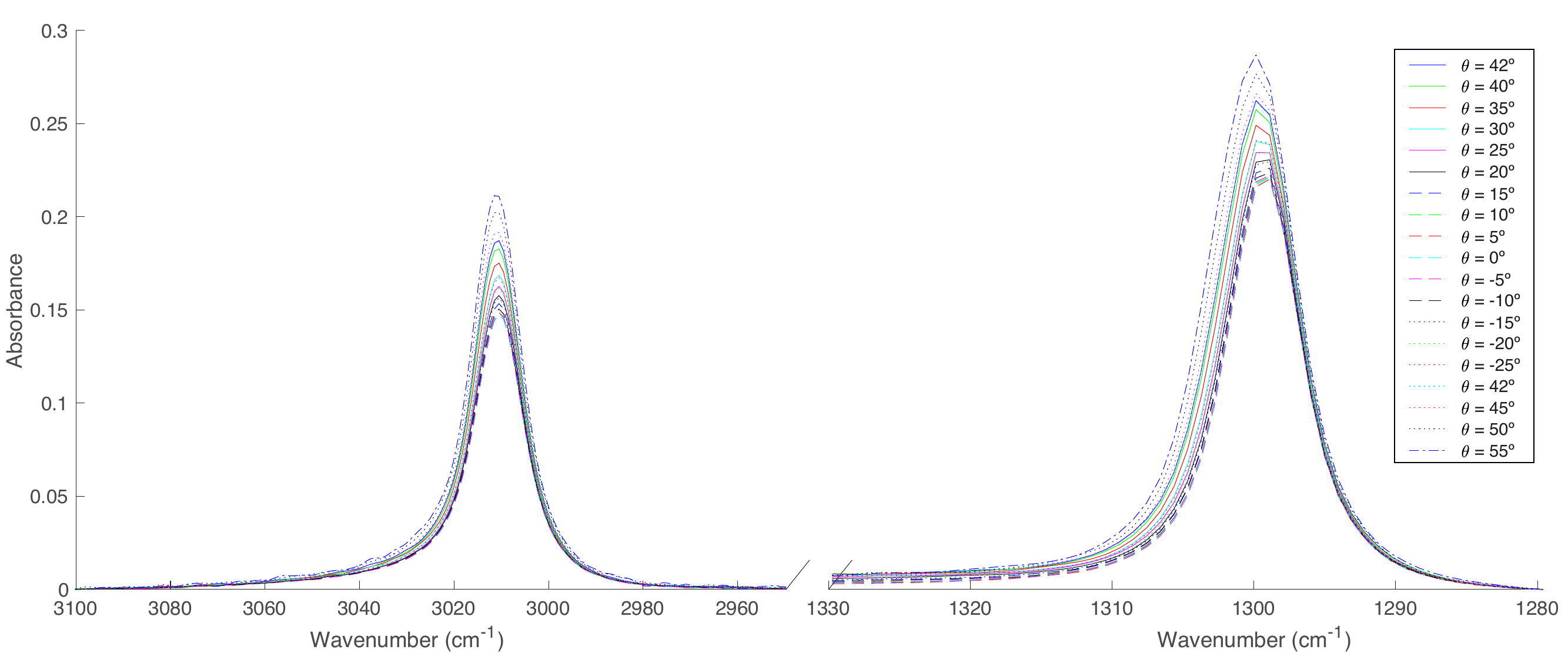}
    \caption{IR spectra of CH$_4$ ice deposited at 11 K, measured at different incident angles. The bands near 1300 cm$^{-1}$ and 3010 cm$^{-1}$ are displayed.}
    \label{fig:trap_Ch4_All}
\end{figure}

\begin{figure}
    \centering
    \includegraphics[width=\textwidth]{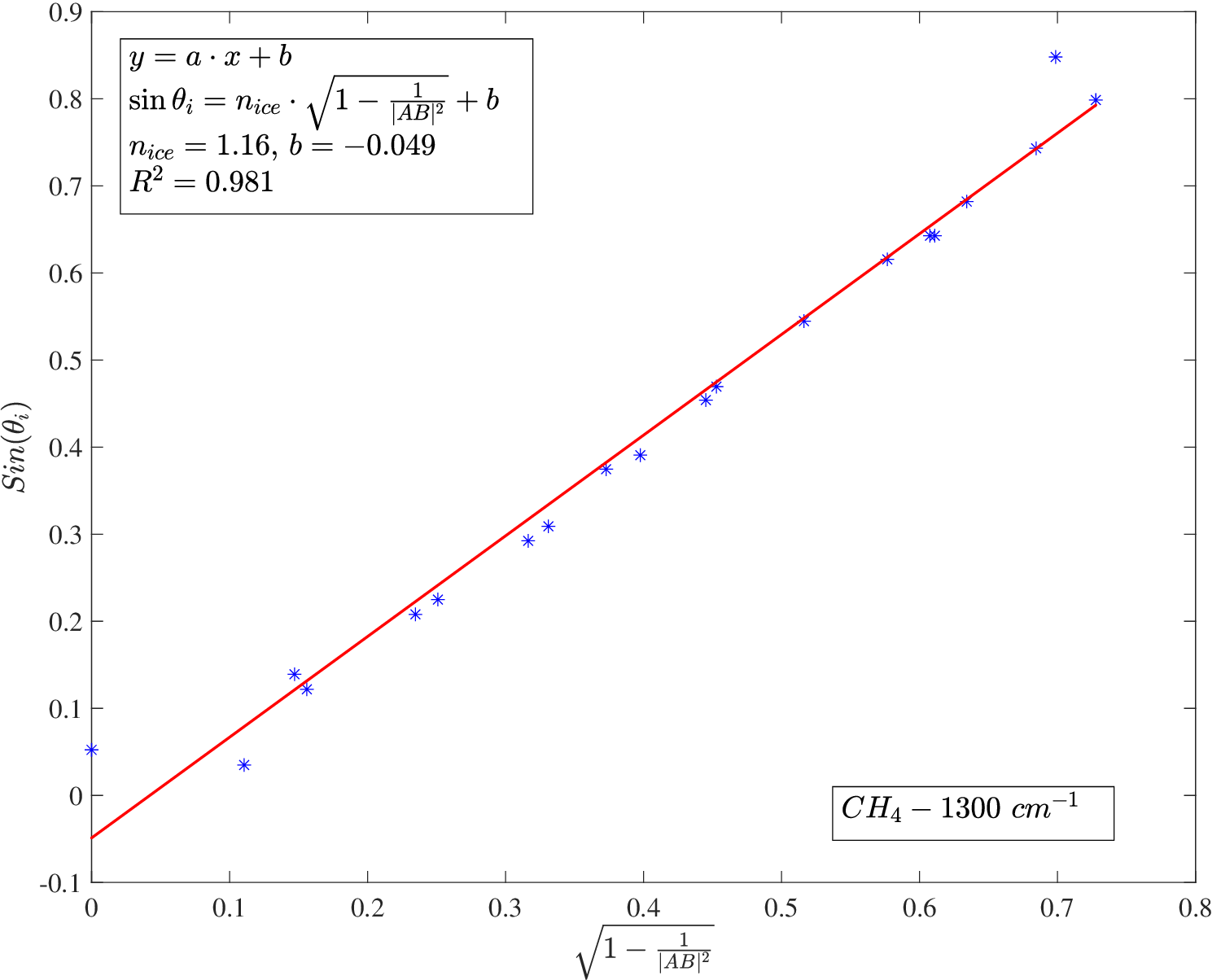}
    \caption{Representation of sine of incident angle of the IR beam, $sin{\theta_i}$, for CH$_4$ ice deposited at 11 K as a function of $\sqrt{1-\frac{d^2}{|AB|^2}}$. Data corresponds to the band around 1300 cm$^{-1}$. The red trace is the linear fit.}
    \label{fig:LinearCH4_1300}
\end{figure}

\begin{figure}
    \centering
    \includegraphics[width=\textwidth]{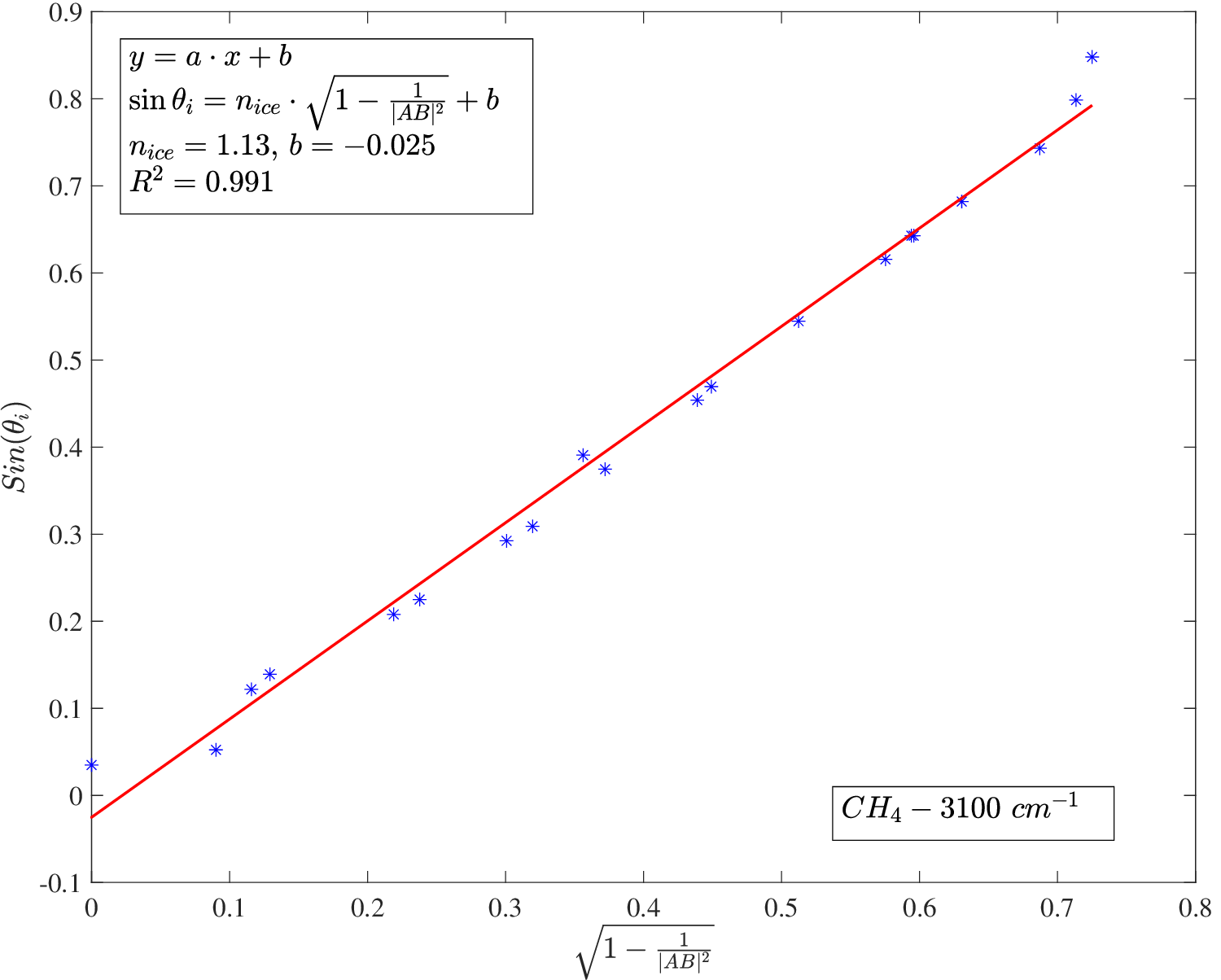}
    \caption{Representation of sine of incident angle of the IR beam, $sin{\theta_i}$, for CH$_4$ ice deposited at 11 K as a function of $\sqrt{1-\frac{d^2}{|AB|^2}}$. Data corresponds to the band around 3100 cm$^{-1}$. The red trace is the linear fit.}
    \label{fig:IR_LinearCH4_3100}
\end{figure}

Fig. \ref{fig:trap_Ch4_All} shows the IR spectra of CH$_4$ ice at different incident angles. 
Figs. \ref{fig:LinearCH4_1300} and \ref{fig:IR_LinearCH4_3100} display $sin{\theta_i}$ as a function of the $\sqrt{1-\frac{d^2}{|AB|^2}}$ for CH$_4$ ice spectra in the 1300 and 3100 cm$^{-1}$ bands. From these plots, the index of refraction are estimated, $n_{CH_4}$ = 1.16 and 1.13 for the 1300 and 3100 cm$^{-1}$ band of CH$_4$, respectively. The corresponding goodness-of-fit values are $R$ = 0.981 and 0.991. These values represent the {\it effective} refractive index for CH$_4$ ice deposited at 11 K.

\subsection{H$_2$S} 
\label{H2S}

\begin{figure}
    \centering
    \includegraphics[width=\textwidth]{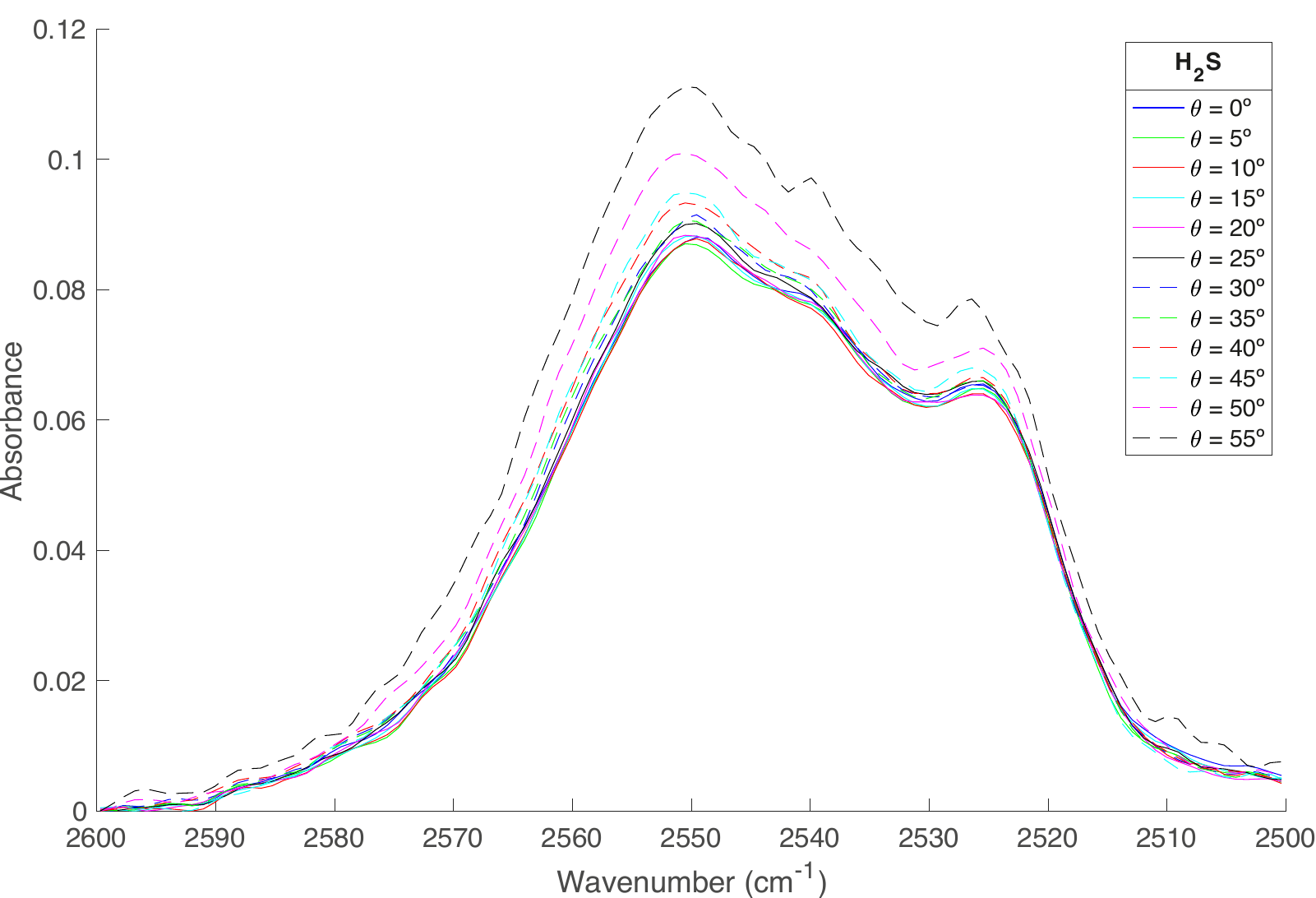}
    \caption{IR band near 2550 cm$^{-1}$ of H$_2$S ice deposited at 11 K as a function of incident angle.}
    \label{fig:IR_h2s_2550cm}
\end{figure}

Fig. \ref{fig:IR_h2s_2550cm} shows the IR spectra of H$_2$S ice at different incident angles. The increase in the angle of incidence that corresponds to a larger optical path across the ice layer, is accompanied by an increase in the IR absorbance. 

\begin{figure}
    \centering
    \includegraphics[width=\textwidth]{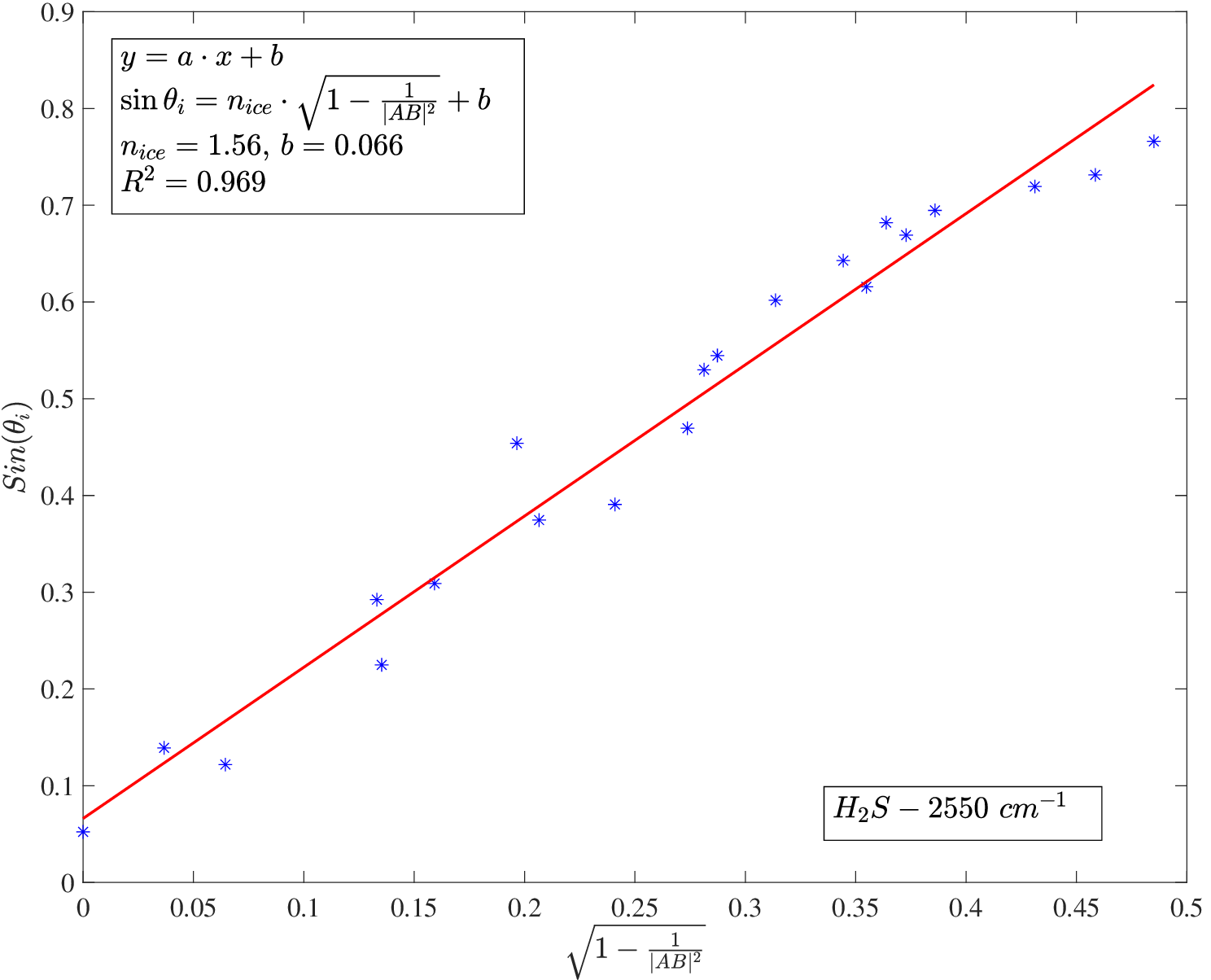}
    \caption{Representation of sine of incident angle of the IR beam, $sin{\theta_i}$, for H$_2$S ice deposited at 11 K as a function of $\sqrt{1-\frac{d^2}{|AB|^2}}$. Data corresponds to the band around 2550 cm$^{-1}$. The red trace is the linear fit.}
    \label{fig:LinearH2S}
\end{figure}

Fig. \ref{fig:LinearH2S} presents $sin{\theta_i}$ as a function of the $\sqrt{1-\frac{d^2}{|AB|^2}}$ in the acquired H$_2$S ice spectra and the linear fit of the data. The estimated index of refraction is $n_{H_2S}$ = 1.56 with a goodness of fit of $R$ = 0.969. As in the previous sections, this should be regarded as the  \textit{effective} index of refraction for H$_2$S ice deposited at 11 K. This value applies to the 2550 cm$^{-1}$ band of H$_2$S.

\begin{table}
\centering
\begin{tabular}{*{5}{c}}
  \toprule
  \toprule
  \textbf{Ice Molecule} & \textbf{$\lambda$ (cm$^{-1}$)} & \textbf{Index of Refraction} & \textbf{R-Square} & \textbf{Correction factor (45$^{\circ}$)}\\
  \cmidrule{1-5}
                    CO & 2138 & 1.17 & 0.983 &  0.797\\ 
  \cmidrule{1-5}
  \multirow{3}{*}{CO$_2$} & 2282 & 1.25 & 0.940 & 0.825\\ 
                          & 2364 & 1.47 & 0.853 & 0.88\\
                          & 3708 & 1.13 & 0.925 & 0.877\\ 
  \cmidrule{1-5}
  H$_2$O & 3283 & 1.00 & 0.974 & 0.707\\ 
  \cmidrule{1-5}
  \multirow{3}{*}{CH$_3$OH}  & 1020 & 1.11 & 0.981 & 0.771\\
                             & 2830 & 1.13 & 0.994 & 0.780\\
                             & 3290 & 1.06 & 0.993 & 0.745\\
  \cmidrule{1-5}                            
  \multirow{2}{*}{NH$_3$}     & 1082 & 1.06 & 0.913 & 0.745\\
                              & 3375 & 1.56 & 0.919 & 0.891\\
  \cmidrule{1-5}
  \multirow{2}{*}{CH$_4$}     & 1300 & 1.16 & 0.981 & 0.793\\
                                & 3100 & 1.13 & 0.991 & 0.780\\  
  \cmidrule{1-5}
  \multirow{1}{*}{H$_2$S}     & 2550 & 1.56 & 0.969 & 0.891\\

  \bottomrule
  \bottomrule
\end{tabular}
\centering \caption{Summary of the \textit{effective} index of refraction $n_{ice}$, R-square values of the linear fit applied to obtain $n_{ice}$ as the slope of this fit, and the correction factor  
$\sqrt{1 - \frac{sin^2{\theta_i}}{n_{ice}^2}}$ in Eq. \ref{N_corrected} for $\theta_i$ = 45$^{\circ}$. The values in this table belong to various bands for different ice compositions. In addition to this correction factor that is dependent on the incident angle $\theta_i$ used for IR spectroscopy, the column density value at normal incidence, $N_{0}$, is also dependent on the integrated absorbance of the IR band that leads to $N_{\theta_i}$ in Eq. \ref{N_corrected}. Note that the values of the correction factor obtained at 45$^{\circ}$ differ significantly from the simple geometric correction factor, i.e. cos(45$^{\circ}$)~=~~0.7071, represented by the dotted green line in Fig. \ref{fig:correction_factor}, showing the error introduced in the ice column density estimation of previous works that merely relied on this 0.7071 factor.}
\label{tab:table_Resumen}
\end{table}

\begin{figure}
    \centering
    \includegraphics[width=\textwidth]{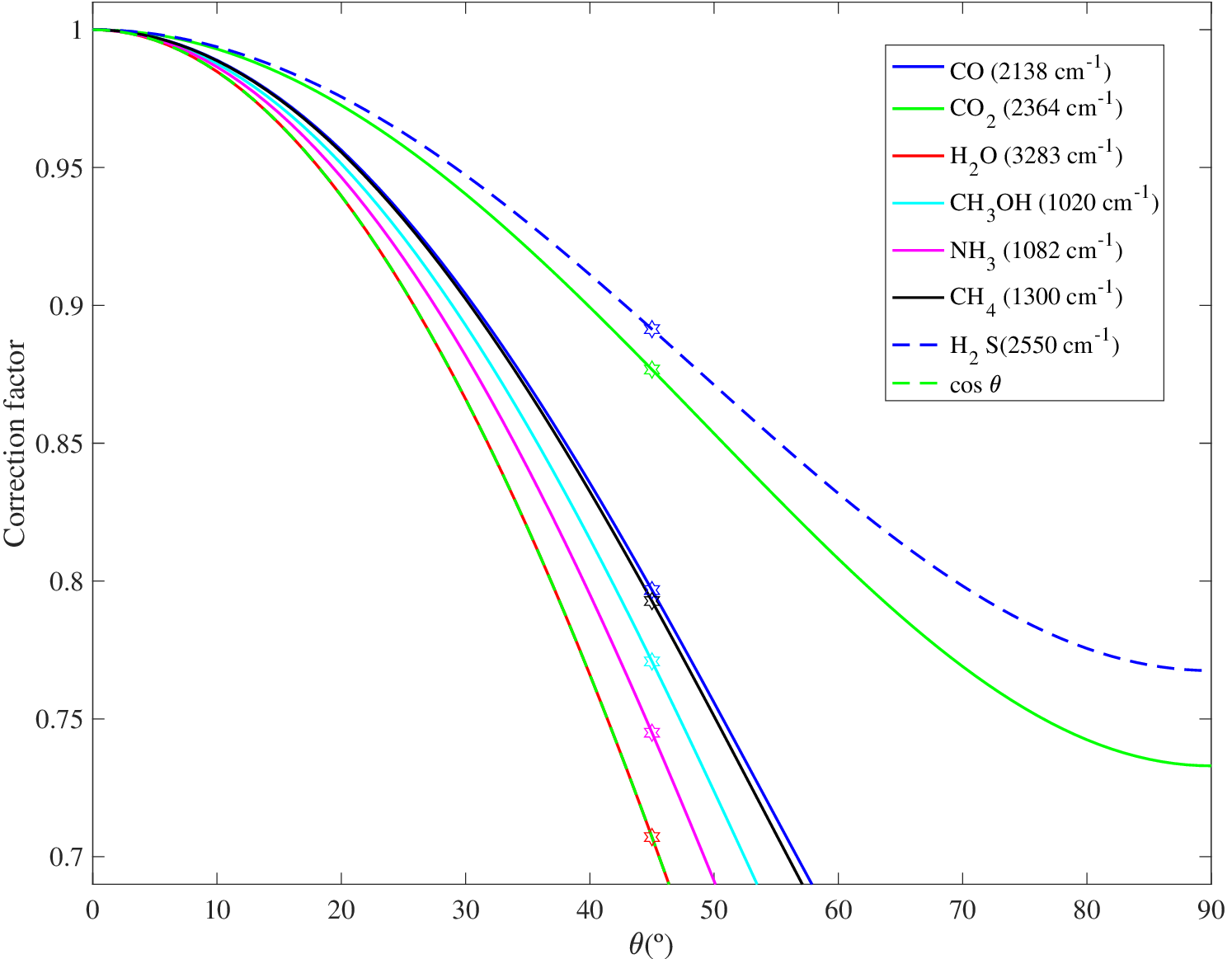}
    \caption{Representation of the correction factor $\sqrt{1 - \frac{sin^2{\theta_i}}{n_{ice}^2}}$ in Eq. \ref{N_corrected} for different incident angle $\theta_i$ values of the IR beam with respect to the ice substrate.}
    \label{fig:correction_factor}
\end{figure}

\section{Conclusions and Astrophysical Implications}
\label{sect:conc}

The {\it effective} index of refraction $n_{ice}$ obtained in this work corresponds to an average value in the spectral range used for the integration of the IR bands. The effective refractive index values presented in Table \ref{tab:table_Resumen} are close to the average value of the refraction index $n$ in the spectral range where these absorption bands are located.
It should be noted that the values of $n_{ice}$ in Table \ref{tab:table_Resumen} are obtained in the infrared. The reader may be more used to the ice values measured in the optical range since they are used to obtain the ice thickness using visible laser interferometry (e.g., \cite{LUNA2022_aa}, \cite{Gonzalez2022}). The values of the correction factor $\sqrt{1 - \frac{sin^2{\theta_i}}{n_{ice}^2}}$ in Eq. \ref{N_corrected} for $\theta_i$ = 45$^{\circ}$ are also presented in Table \ref{tab:table_Resumen}, since this is a commonly used angle in astrochemistry laboratories that perform IR spectroscopy of astrophysical ice analogs. Finally, 
Fig. \ref{fig:correction_factor} is a representation of this correction factor as a function of the incident angle for the various ice compositions studied in this work.  

This paper allows to estimate the correct value of the column density of astrophysically relevant ice components when IR spectroscopy is performed at oblique incidence with respect to the ice sample. More oblique incident angles lead to an increase in the optical path of the IR beam across the ice, which is also dependent on the index of refraction of the ice.
As a result, the IR absorption increases with increasing incident angle. 
In addition, the cases of CO and CO$_2$ ice show the important effect of LO and TO modes on the IR absorption bands, since oblique incident angles intercept both modes. This effect was used to infer information on the ice structure. For instance, a more ordered or crystalline ice structure can be identified because the alignment and orientation of the molecules is reflected in the distribution of LO and TO modes.

The results presented here concern ice layers made of a pure ice component deposited at 11 K. The effect of deposition temperature on the spectroscopy at oblique angles is left for future work. The astrophysical scenario of ice spectroscopy, where the ice mantle build up may take place at an extended temperature range, is much more complex. This ice mantle on top of dust grains will adapt to the morphology of the bare grain surface, which is expected to deviate from a flat surface, and the IR photons will impinge on the ice at different angles. Among other effects, micron-size ice-covered dust grains intercepted along the line of sight suffer from scattering effects at mid-IR wavelengths, this has a considerable impact on the IR band profiles of the ice \citep{Dartois2024}. Furthermore, the interaction of different molecular species in the ice, either with the dominant water ice matrix or among less abundant species, leads to band deformations and wavenumber shifts, or even the formation of new bands that are not present in the pure ice components \citep{MartinDomenech2014}.     

\section*{Acknowledgements}
\label{acknowledgements}
This research has been funded by projects PID2020-118974GB-C21 and PID2023-151513NB-C21 by the Spanish Ministry of Science and Innovation.

\section*{Data Availability}
The data underlying this article will be shared on reasonable request to the corresponding author.

\bibliographystyle{mnras}
\bibliography{GonzalezDiaz} 

\begin{thebibliography}{}
\makeatletter
\relax
\def\mn@urlcharsother{\let\do\@makeother \do\$\do\&\do\#\do\^\do\_\do\%\do\~}
\def\mn@doi{\begingroup\mn@urlcharsother \@ifnextchar [ {\mn@doi@}
  {\mn@doi@[]}}
\def\mn@doi@[#1]#2{\def\@tempa{#1}\ifx\@tempa\@empty \href
  {http://dx.doi.org/#2} {doi:#2}\else \href {http://dx.doi.org/#2} {#1}\fi
  \endgroup}
\def\mn@eprint#1#2{\mn@eprint@#1:#2::\@nil}
\def\mn@eprint@arXiv#1{\href {http://arxiv.org/abs/#1} {{\tt arXiv:#1}}}
\def\mn@eprint@dblp#1{\href {http://dblp.uni-trier.de/rec/bibtex/#1.xml}
  {dblp:#1}}
\def\mn@eprint@#1:#2:#3:#4\@nil{\def\@tempa {#1}\def\@tempb {#2}\def\@tempc
  {#3}\ifx \@tempc \@empty \let \@tempc \@tempb \let \@tempb \@tempa \fi \ifx
  \@tempb \@empty \def\@tempb {arXiv}\fi \@ifundefined
  {mn@eprint@\@tempb}{\@tempb:\@tempc}{\expandafter \expandafter \csname
  mn@eprint@\@tempb\endcsname \expandafter{\@tempc}}}

\bibitem[\protect\citeauthoryear{{Dartois} et~al.,}{{Dartois}
  et~al.}{2024}]{Dartois2024}
{Dartois} E.,  et~al., 2024, \mn@doi [Nature Astronomy]
  {10.1038/s41550-023-02155-x}, \href
  {https://ui.adsabs.harvard.edu/abs/2024NatAs...8..359D} {8, 359}

\bibitem[\protect\citeauthoryear{{Escribano}, {Munoz Caro}, {Cruz-Diaz},
  {Rodriguez-Lazcano}  \& {Mate}}{{Escribano} et~al.}{2013}]{Escribano2013}
{Escribano} R.~M.,  {Munoz Caro} G.~M.,  {Cruz-Diaz} G.~A.,
  {Rodriguez-Lazcano} Y.,   {Mate} B.,  2013, \mn@doi [Proceedings of the
  National Academy of Science] {10.1073/pnas.1222228110}, \href
  {https://ui.adsabs.harvard.edu/abs/2013PNAS..11012899E} {110, 12899}

\bibitem[\protect\citeauthoryear{Falk}{Falk}{1987}]{Falk1987}
Falk M.,  1987, \mn@doi [The Journal of Chemical Physics] {10.1063/1.452307},
  86, 560

\bibitem[\protect\citeauthoryear{{Gonz{\'a}lez D{\'iaz, C} and {Carrascosa, H}
  and {Mu{\~n}oz Caro, G M} and {Satorre, M {\'A}} and {Chen,
  Y-J}}}{{Gonz{\'a}lez D{\'iaz, C} and {Carrascosa, H} and {Mu{\~n}oz Caro, G
  M} and {Satorre, M {\'A}} and {Chen, Y-J}}}{2022}]{Gonzalez2022}
{Gonz{\'a}lez D{\'iaz, C} and {Carrascosa, H} and {Mu{\~n}oz Caro, G M} and
  {Satorre, M {\'A}} and {Chen, Y-J}} 2022, \mn@doi [Monthly Notices of the
  Royal Astronomical Society] {10.1093/mnras/stac3122}, 517, 5744

\bibitem[\protect\citeauthoryear{{Gonz{\'a}lez D{\'i}az, C.}, {Carrascosa de
  Lucas, H.}, {Aparicio, S.}, {Mu{\~{n}}oz Caro, G. M.}, {Sie, N.-E.}, {Hsiao,
  L.-C.}, {Cazaux, S.}  \& {Chen, Y.-J.}}{{Gonz{\'a}lez D{\'i}az, C.}
  et~al.}{2019}]{GonzalezDiaz2019}
{Gonz{\'a}lez D{\'i}az, C.} {Carrascosa de Lucas, H.} {Aparicio, S.}
  {Mu{\~{n}}oz Caro, G. M.} {Sie, N.-E.} {Hsiao, L.-C.} {Cazaux, S.}  {Chen,
  Y.-J.} 2019, \mn@doi [Monthly Notices of the Royal Astronomical Society]
  {10.1093/mnras/stz1223}, 486, 5519

\bibitem[\protect\citeauthoryear{Itoh, Kasuya  \& Hasegawa}{Itoh
  et~al.}{2009}]{itoh2009}
Itoh Y.,  Kasuya A.,   Hasegawa T.,  2009, The Journal of Physical Chemistry A,
  113, 7810

\bibitem[\protect\citeauthoryear{Lasne, Rosu-Finsen, Cassidy, McCoustra  \&
  Field}{Lasne et~al.}{2015}]{Lasne2015}
Lasne J.,  Rosu-Finsen A.,  Cassidy A.,  McCoustra M. R.~S.,   Field D.,  2015,
  \mn@doi [Phys. Chem. Chem. Phys.] {10.1039/C5CP04536C}, 17, 30177

\bibitem[\protect\citeauthoryear{Luna, Millán, Domingo, Santonja  \&
  Satorre}{Luna et~al.}{2022}]{LUNA2022_aa}
Luna R.,  Millán C.,  Domingo M.,  Santonja C.,   Satorre M.~{\'A}.,  2022,
  \mn@doi [The Astrophysical Journal] {10.3847/1538-4357/ac8001}, 935, 134

\bibitem[\protect\citeauthoryear{{Mart{\'\i}n-Dom{\'e}nech}, {Mu{\~n}oz Caro},
  {Bueno}  \& {Goesmann}}{{Mart{\'\i}n-Dom{\'e}nech}
  et~al.}{2014}]{MartinDomenech2014}
{Mart{\'\i}n-Dom{\'e}nech} R.,  {Mu{\~n}oz Caro} G.~M.,  {Bueno} J.,
  {Goesmann} F.,  2014, \mn@doi [\aap] {10.1051/0004-6361/201322824}, \href
  {https://ui.adsabs.harvard.edu/\#abs/2014A&A...564A...8M} {564, A8}

\bibitem[\protect\citeauthoryear{{Mu{\~n}oz Caro}, {Jim{\'e}nez-Escobar},
  {Mart{\'\i}n-Gago}, {Rogero}, {Atienza}, {Puertas}, {Sobrado}  \&
  {Torres-Redondo}}{{Mu{\~n}oz Caro} et~al.}{2010}]{MunozCaro2010}
{Mu{\~n}oz Caro} G.~M.,  {Jim{\'e}nez-Escobar} A.,  {Mart{\'\i}n-Gago}
  J.~{\'A}.,  {Rogero} C.,  {Atienza} C.,  {Puertas} S.,  {Sobrado} J.~M.,
  {Torres-Redondo} J.,  2010, \mn@doi [\aap] {10.1051/0004-6361/200912462},
  \href {https://ui.adsabs.harvard.edu/\#abs/2010A&A...522A.108M} {522, A108}

\bibitem[\protect\citeauthoryear{{Mu{\~n}oz Caro}, {Chen}, {Aparicio},
  {Jim{\'e}nez-Escobar}, {Rosu-Finsen}, {Lasne}  \& {McCoustra}}{{Mu{\~n}oz
  Caro} et~al.}{2016}]{MunozCaro2016}
{Mu{\~n}oz Caro} G.~M.,  {Chen} Y.-J.,  {Aparicio} S.,  {Jim{\'e}nez-Escobar}
  A.,  {Rosu-Finsen} A.,  {Lasne} J.,   {McCoustra} M.~R.~S.,  2016, \mn@doi
  [\aap] {10.1051/0004-6361/201628121}, \href
  {https://ui.adsabs.harvard.edu/\#abs/2016A&A...589A..19M} {589, A19}

\bibitem[\protect\citeauthoryear{\"{O}berg, Facchini  \& Anderson}{\"{O}berg
  et~al.}{2023}]{Obert2011}
\"{O}berg K.~I.,  Facchini S.,   Anderson D.~E.,  2023, \mn@doi [Annual Review
  of Astronomy and Astrophysics] {10.1146/annurev-astro-022823-040820}, 61, 287

\bibitem[\protect\citeauthoryear{{Palumbo}}{{Palumbo}}{2006}]{Palumbo2006}
{Palumbo} M.~E.,  2006, \mn@doi [\aap] {10.1051/0004-6361:20042382}, \href
  {https://ui.adsabs.harvard.edu/abs/2006A&A...453..903P} {453, 903}

\bibitem[\protect\citeauthoryear{{Schiltz, L.} et~al.,}{{Schiltz, L.}
  et~al.}{2024}]{Schiltz_2024}
{Schiltz, L.} et~al., 2024, \mn@doi [A&A] {10.1051/0004-6361/202449846}, 688,
  A155

\bibitem[\protect\citeauthoryear{Zumofen}{Zumofen}{1978}]{Zumofen1978}
Zumofen G.,  1978, \mn@doi [The Journal of Chemical Physics]
  {10.1063/1.436233}, 68, 3747

\makeatother
\end{thebibliography}

\bsp	
\label{lastpage}
\end{document}